# SUMMER INTERNSHIP PROGRAMME 2024

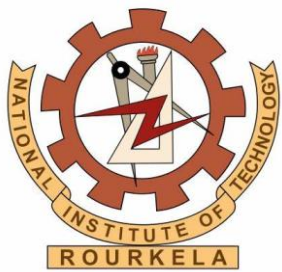 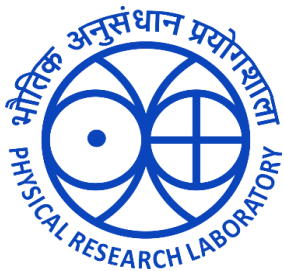 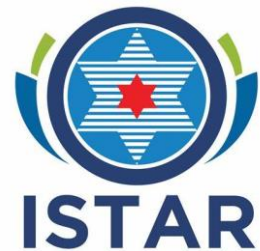

A Project Report

On

## ATMOSPHERIC CHANGES DURING NATURAL DISASTERS: CASE STUDY OF A DUST STORM AND A VOLCANIC ERUPTION USING SPACE AND GROUND-BASED INSTRUMENTS

Submitted by

**DEEPAK KUMAR KAR (421PH5021)**

INTEGRATED MSC. IN PHYSICS AND ASTRONOMY

(NIT ROURKELA)

&

**CHIRAG MAHIDA (23PHY01)**

MASTER OF SCIENCE IN PHYSICS

(ISTAR, CVM UNIVERSITY)

Under the supervision of

**Prof. SOM KUMAR SHARMA**

SPACE AND ATMOSPHERIC SCIENCES DIVISION

# Certificate

# Acknowledgment

The summer internship at the Physical Research Laboratory (PRL), Ahmedabad, under the Department of Space and Atmospheric Sciences, provided a unique and invaluable learning opportunity. This experience has not only broadened our academic horizons but also allowed us to apply theoretical knowledge to practical research, fostering both personal and professional growth.

We would like to express our deepest appreciation to all those who provided us with the possibility to successfully carry out this research project. First and foremost, we extend our heartfelt thanks to **Prof. Som Kumar Sharma** under whom we endeavored this project, and **Mr. Aniket Patel** and **Mr. Dharmendra Kamat**, who helped us through every step of the project. Their expertise and insights were crucial to the successful completion of this work.

We are profoundly grateful to the **Physical Research Laboratory (PRL), Ahmedabad** for allowing us to utilize high-tech instruments and providing extensive resources and support by the PRL Library. We have been fortunate enough to work in the Space and Atmospheric Sciences Division at PRL.

A special thank you goes to NASA Earth Data, the MODIS (Moderate Resolution Imaging Spectroradiometer) team, the SABER (Sounding of the Atmosphere using Broadband Emission Radiometry) mission team, and other open-source satellite organizations. Their provision of satellite data was indispensable to our research. The high-quality data and tools they made available significantly enhanced the scope and accuracy of our project.

We are also thankful to our colleagues and friends at PRL, who offered their time, support, and expertise, making the process smoother and more enjoyable.

We are grateful to the **Director** and **Dean** of Physical Research Laboratory for giving us the opportunity to learn a great deal at this esteemed institution. We would like to express our gratitude to the library staff as well as other PRL staff members who helped us with our research at PRL.

To anyone we may have inadvertently left out, please know that your support and contributions were greatly appreciated.

Name of the students :   Deepak Kumar Kar

                                      Chirag Mahida

Place                  :   PHYSICAL RESEARCH LABORATORY

                                      Navrangpura, Ahmedabad - 380009

# Abstract

This study investigates atmospheric changes during natural disasters, focusing on case studies of a dust storm in Ahmedabad and a volcanic eruption at Mount Ruang. Using the MICROTOPS-II sunphotometer from May 15 to June 19, 2024, the Ozone column, water column height, and Aerosol Optical Thickness (AOT) were measured. Ozone levels followed a diurnal cycle, peaking in the afternoon due to vehicular emissions and increased solar radiation, while water column depth rose with temperature, reflecting higher humidity and evaporation. AOT values increased due to boundary layer dynamics and urban emissions, peaking during rush hours and cloudy conditions.

The dust storm in Ahmedabad on May 13, 2024, highlighted the influence of seasonal variations on dust storm frequency and their impact on atmospheric stability. Ceilometer backscatter data revealed significant boundary layer disruptions and enhanced aerosol mixing, with two obstruction layers identified at 100m and 2500m due to dust and clouds respectively. OMI-Aura satellite data showed a decrease in longwave radiation flux, aligning with observed cooling and increased humidity. The volcanic eruption at Mount Ruang on April 17, 2024, released substantial tephra and aerosols, enhancing cloud formation and decreasing surface temperature. Landsat-9 OLI-II data indicated significant changes in vegetation and cloud cover post-eruption, while OMI-Aura captured elevated SO2 levels as well as an 8.5% decrease in the ozone mixing ratio due to chlorine emissions.

Our findings underscore the critical role of aerosols in climate modulation, acting as cloud condensation nuclei and influencing longwave radiation flux. We utilized a combination of ground-based instruments, such as MICROTOPS II sun photometer, Vaisala ceilometer, and AWS monitoring, along with space-based remote sensing tools like MODIS, OMI-Aura, Landsat 9, and SABER. Integrating observational data with reanalysis models like ERA-5 proved effective for accurate and reliable atmospheric studies. This research highlights the importance of a multi-instrument approach for monitoring and analyzing rapid and complex atmospheric changes, particularly during natural disasters.

**Keywords:** Atmospheric changes, Dust storm, Volcanic eruption, Aerosol Optical Thickness (AOT), Ozone column, Boundary layer dynamics, Longwave flux, Remote sensing

# Table of Contents

# List of Figures

# 1 CHAPTER 1

## 1.1 INTRODUCTION:

### 1.1.1 Earth's Atmosphere:

The solar system is believed to have condensed out of an interstellar gas and dust cloud called 'Primordial Solar Nebula' about 4.6 billion years ago. The atmosphere of Earth along with other terrestrial planets is thought to have formed due to trapping of volatile compounds of the planets itself. The atmosphere is not a closed system, because it exchanges all three of these internally conservative quantities across the atmosphere's boundary below and receives input from regions outside it (Singh, 1995). The atmosphere is the layer of air comprising various constituents surrounding our Earth. The early atmosphere of Earth was a mixture of carbon dioxide ($CO_2$), nitrogen ($N_2$), and water vapor ($H_2O$), with a trace amount of hydrogen ($H_2$) (Seinfeld & Pandis, 2006). The atmosphere thus formed had high hydrogen and little or no oxygen, so could not support complex life forms. However, as the earth cooled down further, complex chemical actions and reactions in the crust and the interactions between the crust and the atmosphere gradually led to the formation of an atmosphere that could support an early form of life such as single-celled microbes that required little oxygen for their survival (Saha, 2008). Gases, clouds, rain, snow crystals, etc. all are a part of the Earth's atmosphere and they dictate the weather as well as the climate of a specific geographical location controlled by the variation in temperature, pressure, and density profiles with altitude. With our models and data gathered from various remote sensing techniques, we now have a better understanding of the atmosphere to predict accurate weather forecasts and other meteorological phenomena (Wallace & Hobbs, 2006).

### 1.1.2 Composition:

The gases of the Atmosphere are always in motion due to the variations in temperature, pressure, and gravitational effects of the Earth rotating. The Earth's atmosphere is composed primarily of the gases $N_2$ (78%), $O_2$ (21%), and Ar (1%), whose abundances are controlled over geologic timescales by the biosphere, uptake and release from crustal material, and degassing of the interior (Seinfeld & Pandis, 2006) . Minor constituents at low levels of the atmosphere may include variable quantities of dust, smoke, and toxic gases and vapors such as carbon dioxide, sulphur dioxide, methane, oxides of nitrogen, etc., some of which pollute the atmosphere and are highly injurious to health (Saha, 2008) . These gaseous constituents, the trace gases, represent less than 1% of the atmosphere. These trace gases play a crucial role in the Earth's radiative balance and in the chemical properties of the atmosphere. The trace gas abundances have changed rapidly and remarkably over the last two centuries (Seinfeld & Pandis, 2006). Water vapor, although variable in concentration, is also a significant component, especially in the lower layers where it is critical for weather and climate processes (Jacob, 1999). The concentration of Ozone ($O_3$) also varies a lot throughout a day. Exposure to Ozone concentration >0.1ppmv is considered hazardous to human health (Wallace & Hobbs, 2006).

Infrared (IR) active gases, principally water vapor ($H_2O$), carbon dioxide ($CO_2$), and ozone ($O_3$), naturally present in the Earth's atmosphere, absorb thermal IR radiation emitted by the Earth's surface and atmosphere. The atmosphere is warmed by this mechanism and, in turn, emits IR radiation, with a significant portion of this energy acting to warm the surface and the lower atmosphere. As a consequence, the average surface air temperature of the Earth is about 30°C higher than it would be without atmospheric absorption and reradiation of IR energy (Kellogg, 1996; Peixoto and Oort, 1992; Henderson-Sellers and Robinson, 1986). This phenomenon is popularly known as the "greenhouse effect", and the IR active gases responsible for the effect are likewise referred to as "greenhouse gases" (Ledley, 1999).

*Table 1. Mixing ratios of gases in dry air*

| Gas | Mixing Ratio (mol/mol) |
|---|---|
| Nitrogen ($N_2$) | 0.78 |
| Oxygen ($O_2$) | 0.21 |
| Argon (Ar) | 0.0093 |
| Carbon dioxide ($CO_2$) | $365 \times 10^{-6}$ |
| Neon (Ne) | $18 \times 10^{-6}$ |
| Ozone ($O_3$) | $0.01\text{-}10 \times 10^{-6}$ |
| Helium (He) | $5.2 \times 10^{-6}$ |
| Methane ($CH_4$) | $1.7 \times 10^{-6}$ |
| Krypton (Kr) | $1.1 \times 10^{-6}$ |
| Hydrogen ($H_2$) | $500 \times 10^{-9}$ |
| Nitrous oxide ($N_2O$) | $320 \times 10^{-9}$ |

### 1.1.3 Layers:

We can divide the atmosphere into concentric shells around the Earth. Each shell varies in thickness, density, thermodynamical parameters like temperature, pressure, and heat (Gradwell, 2006)

The atmosphere is conventionally divided into layers in the vertical direction, according to the variation of temperature with height. The layer from the ground up to about 15 km altitude, in which the temperature decreases with height, is called the troposphere and is bounded above by the tropopause. The layer from the tropopause to about 50 km altitude, in which the temperature rises with altitude, is called the stratosphere and is bounded above by the stratopause. The layer from the stratopause to about 85–90 km, in which the temperature again falls with altitude, is called the mesosphere and is bounded above by the mesopause. Above the mesopause is the thermosphere, in which the temperature again rises with altitude (Andrews, 2010.).

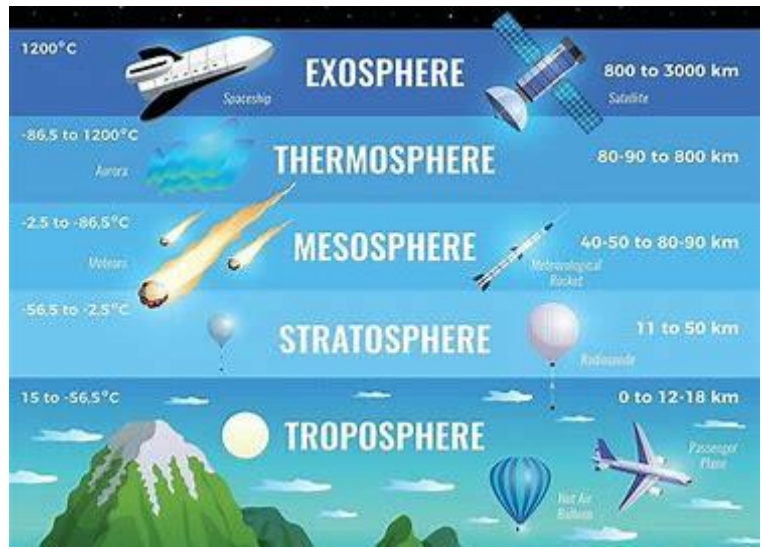

*Figure 1. Layers of the atmosphere. (Source: www.worldatlas.com/articles/what-are-the-5-layers-of-the-earth-s-atmosphere.html)*

**Troposphere:**

Troposphere is the lowest layer of the atmosphere, extending from Earth's surface to about 10-15 km till tropopause. It contains about 75% of the atmosphere's mass and the majority of water vapor, clouds, and weather phenomena take place in this region. There is a decrease in temperature with an increase in altitude. This is because the surface of the earth is hotter and temperature decreases as we move further. The decrease in pressure also plays an important role in the positive lapse rate. Maximum air temperature also occurs near the Earth's surface in this layer. With increasing height, the air temperature drops uniformly with altitude at a rate of approximately 6.5° Celsius per 1000 meters. This phenomenon is commonly called the Environmental Lapse Rate. At an average temperature of -56.5° Celsius, the top of the troposphere is reached (Pidwirny, 2006). Vertical mixing of air is rapid due to the temperature gradient.

**Stratosphere:**

Above the tropopause is the stratosphere. This layer extends from an average altitude of 11 to 50 kilometers above the Earth's surface. This stratosphere contains about 19.9% of the total mass found in the atmosphere. Very little weather occurs in the stratosphere. Occasionally, the top portions of thunderstorms breach this layer. The lower portion of the stratosphere is also influenced by the polar jet stream and subtropical jet stream. In the first 9 kilometers of the stratosphere, temperature remains constant with height. A zone with constant temperature in the atmosphere is called an isothermal layer. From an altitude of 20 to 50 kilometers, temperature increases with an increase in altitude. The higher temperatures found in this region of the stratosphere occur because of a localized concentration of ozone gas molecules. These molecules absorb ultraviolet sunlight creating heat energy that warms the stratosphere. Ozone is primarily found in the atmosphere at varying concentrations between the altitudes of 10 to 50 kilometers. This layer of ozone is also called the ozone layer. The ozone layer is important to organisms at the Earth's surface as it protects them from the harmful effects of the Sun's ultraviolet radiation. Without the ozone layer life could not exist on the Earth's surface.

**Mesosphere:**

The mesosphere extends from 50 to 85 kilometers above the Earth and is marked by decreasing temperatures with altitude, making it the coldest atmospheric layer. The decrease in temperature is due to the rarefied atmosphere with little to no solar absorption. This region is where most meteors burn up upon entering the Earth's atmosphere (Gadsden & Schröder, 1989) gradually transitioning into the vacuum of space. It is the outermost layer, where atmospheric particles are so sparse that they can travel hundreds of kilometers without colliding (Roble & Ridley, 1987).

**Thermosphere:**

The region above the mesopause is characterized by high temperatures as a result of the absorption of short-wavelength radiation by $N_2$ and $O_2$. The thermosphere stretches from approximately 85 to 600 kilometers and experiences a dramatic increase in temperature with altitude This layer has a rapid vertical mixing. The ionosphere is a region of the upper mesosphere and lower thermosphere where ions are produced by photoionization (Seinfeld & Pandis, 2006). The ionosphere is essential for radio communication as it reflects and modifies radio waves (Kivelson & Russell, 1995). Radiation causes the atmospheric particles in this layer to become electrically charged, enabling radio waves to be refracted and thus be received beyond the horizon.

**Exosphere:**

The main components of the earth's exosphere are neutral oxygen, neutral hydrogen atoms, ionized oxygen, and ionized hydrogen. The position of the base of the exosphere is established at an altitude of 530 km, from an analysis of density data obtained from satellite drag observations. The relative distribution of both neutral oxygen and neutral hydrogen is derived from a theory of the exosphere (Singer, 1960). As the distance from Earth to the exosphere is greater than 500km, the gas molecules with sufficient thermal energy can escape the Earth's gravitational pull.

### 1.1.4 Temperature Variation with Altitude:

Temperature gives a measure of heat in a body. When a body is heated or cooled, its temperature changes. However, given the same quantity of heat, the rise in temperature is not quite the same for all bodies. It depends on a property of the body called its heat capacity which is given by the product of its mass and a quantity called specific heat. Specific heat of a body is defined as the quantity of heat required to raise the temperature of the unit mass of the body through 1◦C. Thus, specific heat is related to a given quantity of heat by the relation:

Heat added (or subtracted) = mass × specific heat × rise (or fall) in temperature

The unit of heat is a calorie which is the quantity of heat required to raise the temperature of 1 g of water through 1◦C, usually from 15 to 16◦C. Heat is measured by calorimeters.

French meteorologist Leon Philippe Teisserenc de Bort. Sending up temperature-measuring devices in balloons, he found that, contrary to the popular belief of the day, the temperature in the atmosphere did not steadily decrease to absolute zero with increasing altitude, but stopped falling and remained constant at 11 km or so. After that meteorologists started to map the variation of temperature with altitude.

Observations from Radiosonde, weather radars, lidar, and various remote sensing methods give an accurate description of the temperature profile of the Earth's atmosphere.

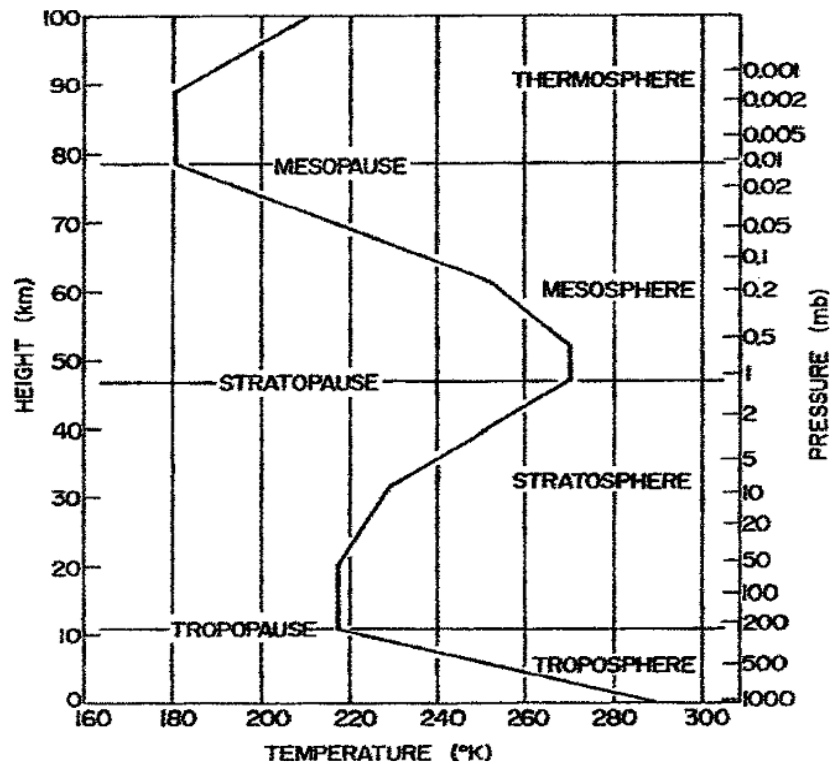

*Figure 2. Vertical distribution of temperature in the U.S. standard atmosphere (Seinfeld & Pandis, 2006)*

In the figure 2, we can see the variation of temperature (in K) with the altitude (in km). The temperature decreases from the surface as we go up, till we reach a level called tropopause. By convention of the World Meteorological Organization (WMO), the tropopause is defined as the lowest level at which the rate of decrease of temperature with height (the temperature lapse rate) decreases to 2K km$^{-1}$ or less and the lapse rate averaged between this level and any level within the next 2 km does not exceed 2K km$^{-1}$ (Holton, et al., 1995).

In stratosphere, temperature generally increases linearly up to stratopause, but there is an isothermal region from 11 km to about 20 km where temperature does not vary much. This is because, in the lower stratosphere, there is a balance between the radiative heating from the absorption of UV radiation by ozone and the radiative cooling due to infrared emissions from the atmosphere. This balance results in a relatively stable temperature profile, creating this isothermal region. Essentially, the amount of heat absorbed from the sun and the heat lost through radiation are roughly equal, leading to little temperature variation (Wallace & Hobbs, 2006).

The stratosphere ends with stratopause, after which temperature starts decreasing in mesosphere. The temperature reaches a minimum of -100°C, the coldest point of the atmosphere, called mesopause. After this layer, the temperature keeps on increasing in a linear manner. This uppermost layer is called the thermosphere. The importance of this layer lies in the fact that it intercepts the highly-charged solar rays from space and the high-energy ultra-violet radiation from the sun which are both harmful to life at the earth's surface. The atoms and molecules of gases such as oxygen and nitrogen present in this layer absorb the high-energy short-wave radiation from the sun and get ionized. The ionized layer by reflecting radio waves helps in global telecommunication. For this reason, this layer is also sometimes called the ionosphere (Saha, 2008).

As for the variation of altitude effect with seasonal change and global warming, we think that the altitude is not only a simple altitude effect but also a topography effect (Aigang, Tianming, Shichang, & Deqian, 2009 ). The temperature profile also depends on many factors like solar activity. The temperature of the thermosphere varies greatly with solar activity, with a value of about 2000K at the time of 'active sun' and 500K at the time of 'quiet sun' at 500 km altitude (Banks & Kockarts, 2013).

### 1.1.5  Pressure and Density Variation with Height:

The unit of pressure in the International System of Units (SI) is newtons per meter squared (N m$^{-2}$), which is called the pascal (Pa). In terms of pascals, the atmospheric pressure at the surface of the Earth, the so-called standard atmosphere, is $1.01325 \times 10^5$ Pa. Another commonly used unit of pressure in atmospheric science is the millibar (mbar), which is equivalent to the hectopascal (hPa). The standard atmosphere is 1013.25 mbar.

To derive an expression for pressure variation with height, we need to introduce the concept of "Hydrostatic Equilibrium".

For an atmosphere at rest, in static equilibrium, the net forces acting on any small portion of air must balance. Consider for example a small cylinder of air, of height Δz and horizontal cross-sectional area ΔA. This is subject to a gravitational force gΔm downwards, where its mass Δm = ρ ΔAΔz and g is the gravitational acceleration (assumed constant). This force must be balanced by the difference between the upward pressure force p(z) ΔA on the bottom of the cylinder and the downward pressure force p(z + Δz) ΔA on the top. We therefore have:

$$g\rho \, \Delta A \, \Delta z = \big(p(z) - p(z + \Delta z)\big)\Delta A \;;$$

by canceling ΔA and using the Taylor expansion;

$$p(z + \Delta z) \approx p(z) + \frac{dp}{dz}\Delta z$$

we get the equation for hydrostatic balance,

$$\frac{dp}{dz} = -\rho g$$

We can derive some basic properties of the atmosphere, given that it is an ideal gas and assuming that it is in hydrostatic balance,

$$\frac{d}{dz}(\ln p) = -\frac{g}{R_a T}$$

Where;

p(z) = pressure at height z,

$R_a$ = universal gas constant,

T = temperature profile (can be taken isothermal = $T_0$)

Integrating it, we get;

$$p = p_0 e^{\frac{-gz}{R_a T_0}} = p_0 e^{-\frac{z}{H}}$$

*(1)*

where H = $R_a T_0/g$ is the pressure scale height, the height over which the pressure falls by a factor of e. In this isothermal case the density also falls exponentially with height in the same way: ρ = ρ$_0$ exp(−z/H), ρ$_0$ being the density at the ground. For an isothermal atmosphere with T= 260K, H is about 7.6 km.

The lapse rate Γ denotes the rate of decrease of temperature with height:

$$\Gamma(z) = -\frac{dT}{dz};$$

*(2)*

In general, Γ > 0 (decreases with height) in troposphere and Γ < 0 in stratosphere. A layer with Γ < 0 is called inversion layer.

We can actually calculate the number density of air molecules as a function of altitude by using the scale height H = 7.4 km and using

$$n_{air}(z) = n_{air}(0)e^{-\frac{z}{H}};$$

*(3)*

where n$_{air}$ (0) is the number density at the surface.

*Table 2. U.S. Standard Atmosphere 1976 (Seinfeld & Pandis, 2006)*

| z (km) | $n_{air}$ (molecules cm$^{-3}$) Approximate | U.S. Standard Atmosphere$^a$ |
|---|---|---|
| 0 | 2.55 × 10$^{19}$ | 2.55 × 10$^{19}$ |
| 5 | 1.3 × 10$^{19}$ | 1.36 × 10$^{19}$ |
| 10 | 6.6 × 10$^{18}$ | 6.7 × 10$^{18}$ |
| 15 | 3.4 × 10$^{18}$ | 3.0 × 10$^{18}$ |
| 20 | 1.7 × 10$^{18}$ | 1.4 × 10$^{18}$ |
| 25 | 8.7 × 10$^{17}$ | 6.4 × 10$^{17}$ |

### 1.1.6 Ozone:

The ozone ($O_3$) is the tri-molecular oxygen compound found at its maximum concentration in the stratosphere. The amount of ozone required to shield the earth's surface is believed to exist since 600 million years ago. At this time the oxygen concentration was about 10% of what it is present today. The presence of ozone enabled the organisms to develop and live on the land. Before this, there were few organisms who thrive in the water itself.

**Ozone formation**

The ozone formed with the photochemical reaction in the atmosphere. The oxygen was liberated by the process of photosynthesis by the green plants. The ultraviolet rays are absorbed by the oxygen molecules. The radiation below the wavelength of 240 nm has sufficient energy to dissociate the oxygen molecule. The cloudless environment and warm conditions at high-pressure systems are favorable for the formation of ozone molecules.

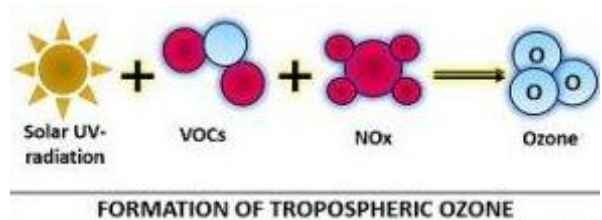

FORMATION OF TROPOSPHERIC OZONE

The cloudless environments and warm conditions at high-pressure systems are favorable for the formation of ozone molecules.

$$O_2 + hv \rightarrow O + O$$

$$O_2 \rightarrow 2O$$

The oxygen again combines with the molecular oxygen present and leads to the formation of ozone as in the following reaction: -

$$O_2 + O + M = O_3 + M^*$$

Here M is the third body that absorbs the excess energy liberated during the combination of O and $O_2$.

UV radiation can impair the development of cell-mediated immunity in human beings. (Cooper, 1992) Irradiation of skin with ultraviolet B (UVB), the sunburning part of the solar spectrum at the earth's surface, inhibits the development of contact sensitivity to dinitrochlorobenzene applied to the irradiated. UVB may also inhibit sensitization to an allergen applied to non-irradiated skin; thus, the effects of UVB on cell-mediated immunity may be systemic as well as local.

The potential importance of these effects is evidence that UVB can impair the development of contact sensitization in people with black skin as well as people with white skin. (Vermeer, 1991) Thus any clinical effects of UVB suppression of cell-mediated immunity may be applicable to a much larger proportion of the world's population than skin cancer, for example.

Incident UV radiation from the sun must first run the gauntlet of backscattering into space, absorption in the atmosphere, and absorption on the ground before it can strike a human being and produce some effect. (Frederick, 1988)

UV radiation is divided between three wavelength bands: UVC, 100 to 280 nm, UVB, 280-315 nm and UVA, 315-400 nm. Visible light lies in the 400-780 nm wavelength band. Because of stratospheric ozone, only radiation in the UVB and UVA bands of UV radiation reaches the surface of the earth. The bulk of harm to human health probably comes from UVB. This is the band most strongly influenced in intensity by the amount of atmospheric ozone.

### 1.1.7  Water vapour:

Water in atmosphere exists in all three distinct phases i.e. solid, liquid, and gaseous. It mainly enters into the atmosphere through the process of evaporation from the soil, water bodies like streams, lakes, rivers, and ocean surfaces, sublimation from the ice and glaciers, and transpiration from plant leaves. Apart from this the water vapour also comes from combustion, respiration, volcanic eruptions, and other biological and geological process. Its residence time in the atmosphere is about 9 to 10 days. Approximately 85% of the vapor comes from the evaporation of the ocean surface. (Jacobson, 2005)

Water vapor is a principal atmospheric variable, and is a central component in both the earth energy budget and the Global Water Cycle. (Vonder Haar, 2012)water vapor is mainly concentrated in the troposphere i.e. about 99.13 %.

A greenhouse gas, the water vapor feedback roughly doubles the warming induced by $co_2$ increases. It is well known that water vapor generally increases with rising air temperature. The increasing rate of 6% - 7% K−1 for the saturation vapor pressure (Ning, 2016), as the atmospheric relative humidity changes a little (Ming, 2018)For example, the column-integrated atmospheric water vapor has increased with surface temperature at a rate of approximately 7%–9% K −1 for the oceans from 1988–2003 (Trenberth, 2005).

The water vapor increase will affect the hydrological cycle, leading to changes such as large increases in both precipitation and storm intensity and increased moisture convergence in the deep tropics (Dai, 2021)

In addition, near-surface actual water vapor pressure (AVP) strongly regulates surface vapor pressure deficit (VPD), that is, the difference between SVP and AVP. The increased VPD is a key driver of increased aridity and drought under global warming. (Scheff, 2014)

### 1.1.8  Aerosols:

Aerosols are defined as the disperse system with the air as the carrier gas and as a solid or liquid disperse phase or mixture of both. They are the colloidal particles present in the atmosphere. It is the solid or liquid particles that interact with most of the atmospheric phenomenon. Its size ranges from 1 nanometre to several hundred micrometers. They are larger than the atmospheric ions and even as large as the cloud droplets or ice crystals. As compared to the cloud droplets or ice crystals they are homogeneously mixed.

There are different modes of the aerosols according to their diameter such as follows:

1) Nucleation mode: - $10^{-3} \mu m - 10^{-2} \mu m$
2) Aitken mode: - $10^{-2} \mu m - 0.1 \mu m$
3) Accumulation mode: - $0.1 \mu m - 1 \mu m$
4) Coarse mode: - particles >$1 \mu m$

**Types of aerosols:**

There are three types of aerosols
1) Continental aerosol
2) Marine aerosol
3) Stratospheric aerosol

1. Continental aerosol
   The aerosol generated on the land surface is known as the continental aerosol. There are various processes involved in the formation of such aerosols such as the crustal species produced by the erosion of the earth's surface, combustion and secondary components related to anthropogenic activities, carbonaceous components consisting of hydrocarbons volcanic eruptions, etc.

   The crustal aerosols are produced in subtropical deserts like the Sahara, the southwestern USA, and southern Asia. Mainly the aerosols are produced in the northern hemisphere while are found in the southern hemisphere.

2. Marine aerosols
   Aerosols formed above the surface of the oceans or seas are called marine aerosols. They are primarily composed of sea salt having a small over all concentration. The number density of aerosols drops sharply above the boundary layer. As compared to the continental aerosol their number is much more.

3. Stratosphere aerosol
   Aerosols in the stratosphere are introduced through penetrative convection and volcanic eruptions.(Salby, 1996)

## 1.2 ATMOSPHERIC BOUNDARY LAYER:

Atmospheric Boundary Layer (ABL) also called Planetary Boundary Layer (PBL) is a significant layer of our atmosphere to study atmospheric variations and different fields, including air pollution, agricultural meteorology, hydrology, aeronautical meteorology, mesoscale meteorology, weather forecasting and climate (Garratt., 1994).

Atmospheric Boundary Layer is the part of the atmosphere that is directly influenced by the Earth's surface and the phenomena taking place in the Earth's surface. It responds to the surface forcings of Earth with a time scale of an hour or less. The forcing consists of frictional drag, evaporation and transpiration heat transfer, pollutant emission, and terrain-induced flow modification (Stull, 1988).

The thickness of the BL (Boundary Layer) is quite dynamic and varies with time and place. It is normally 1 or 2 km thick (i.e., occupying around the bottom 10 to 20% of the troposphere), and it can range from tens of meters to 4 km or more. Over ocean regions, the depth of boundary layer varies relatively slowly from that of land. Over land area, the boundary layer has a well-defined structure and shows a noticeable diurnal variation (Stull, 2005).

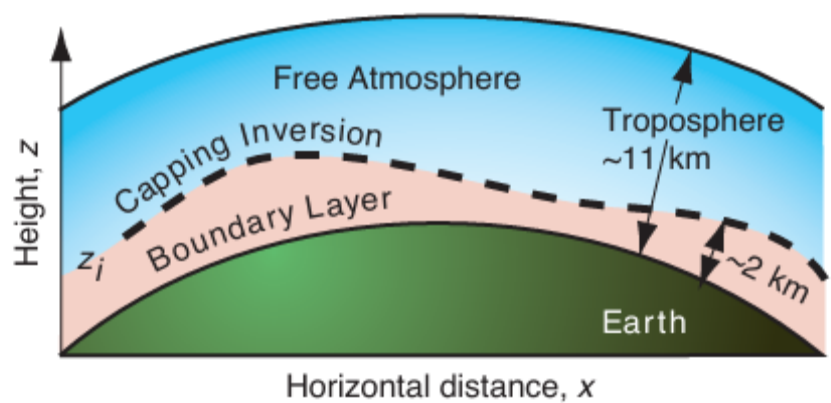

*Figure 3. Atmospheric Boundary Layer and its constituents (Stull, 1988)*

Here we can see the boundary layer varying with the altitude. Due to variations in temperature in the lower Troposphere, the height of the boundary layer changes. This type of variation happens in a cycle every day, known as diurnal variation. Such diurnal variation is one of the key characteristics of the boundary layer over land. There is not much variation near the end of the boundary layer, where the free atmosphere just begins. It is stable at that height and shows little to no diurnal variation.

This diurnal variation is not caused by the direct impact of solar radiation on the ABL. A very low amount of the solar radiation incident is absorbed by the boundary layer while most of it is transmitted to the ground where typical absorptivity is on the order of 90% resulting in the absorption of much of the solar energy. The warming and cooling of the ground in response to the radiation forces changes in the boundary layer via transport processes. Turbulence is one of the important transport processes and is sometimes also used to define the boundary layer (Stull, 1988). These swirls and eddies carry moisture, heat, momentum, and pollutants to regulate the atmosphere.

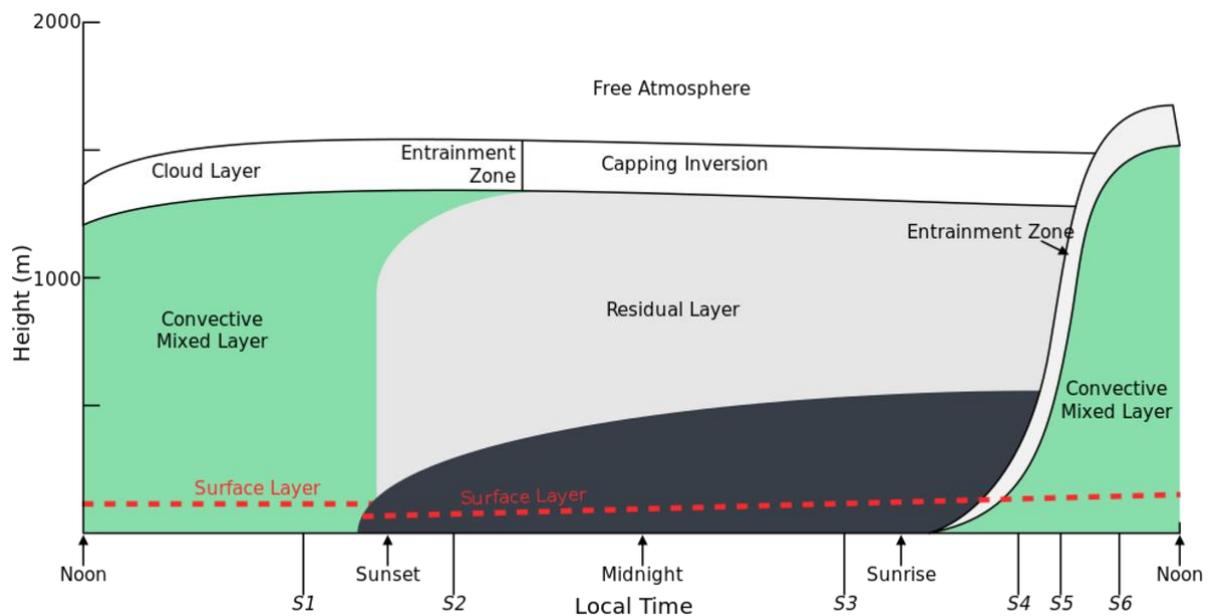

*Figure 4. Diurnal variation of the boundary layer and its types*

*(Source: https://www.e-education.psu.edu/meteo300/node/712)*

ABL can be further divided into various layers that appear and disappear during the diurnal variation cycle. Here we will list out all the different layers that can form during a cycle:

### 1.2.1   Surface Layer:

This is the topmost layer of Earth's crust that is in direct contact with the atmosphere. The solar radiation directly falls on it with very little absorption by the boundary layers. If we imagine the land surface as an infinitesimally thin surface that has zero heat capacity, then the heat flux coming in must balance the heat left to keep the surface energy balance intact. The radiated heat from the surface layer along with the rotation of the Earth initiates the diurnal variation.

### 1.2.2   Capping Inversion:

The top of the boundary layer in convective conditions is often well defined by the existence of a stable layer called capping inversion, into which turbulent motions in the lower layers are generally unable to penetrate very far, though they may continually erode it by heating of gases. The height of this elevated stable layer is quite variable but is generally below 2 to 3 km. The top of a convective boundary layer is well defined in Fig. 5 by the sharp decrease in aerosol concentration at a height of about 1200 m and coincides with the base of a deep and intense subsidence inversion. The capping inversion traps most of the aerosols, turbulence, and moisture below it and prevents it from reaching the free atmosphere.

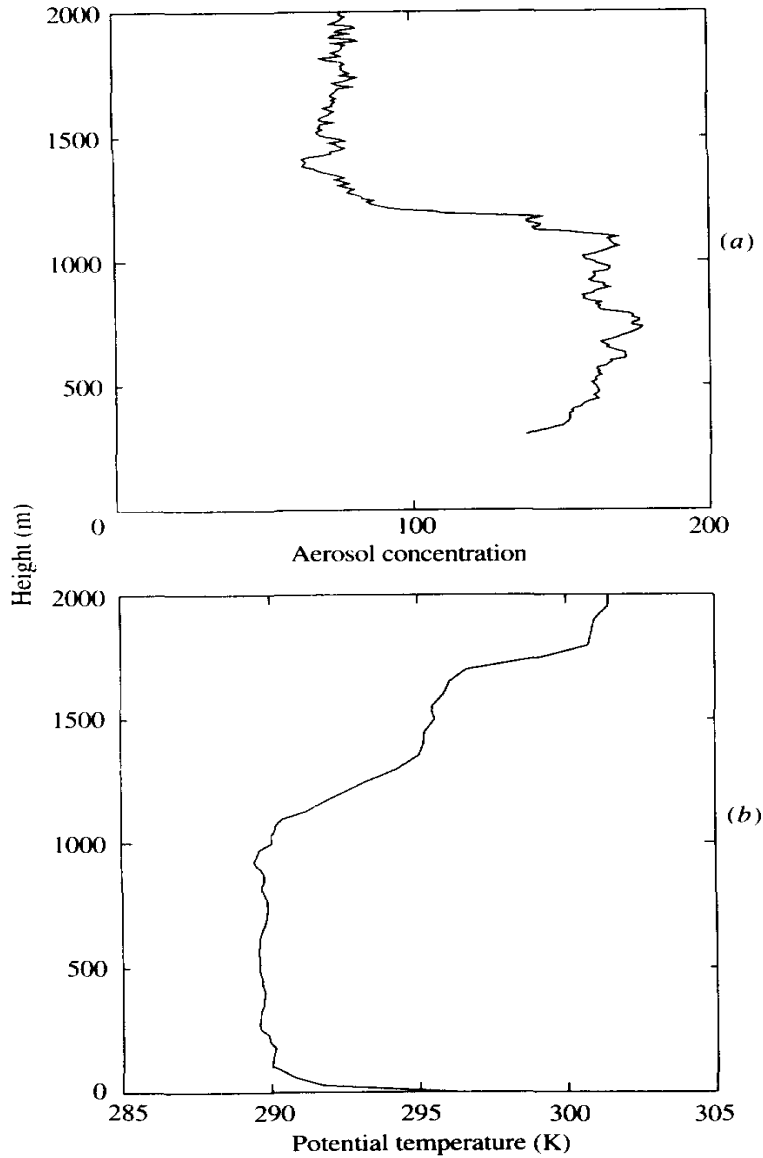

*Figure 5. Vertical profiles of (a) aerosol concentration in arbitrary units and (b) potential temperature observed overland in convective conditions (see Garratt, 1992; fig. 1.2).*

### 1.2.3 Convective Mixed Layer:

Rising thermals and turbulent eddies can overshoot due to their inertia causes a pressure gradient that drives the air from the free atmosphere that causes intense mixing of the air. This creates the mixing layer and is convective in nature. Convective sources include heat transfer from a warm ground surface and radiative cooling from the top of the cloud layer.

In cloud-free weather, the ML growth is proportional to the heating of the ground due to the sun. The formation of the layer starts about a half hour after sunrise, a turbulent ML begins to grow increasing its depth. This ML is characterized by intense mixing in an unstable situation where thermals of warm air rise above the ground. The ML reaches its maximum depth in the late afternoon. It grows by entraining or mixing down into it, the less turbulent air from the free atmosphere above (Stull, 1988).

### 1.2.4 Free Atmosphere:

This is the layer that marks the end of the boundary layer and forms a standard atmosphere having stability. The height of it is not fixed as the height of the boundary layer is dynamic. The free atmosphere extends till the end of troposphere. It is characterized by static stability i.e., buoyancy forces, resist vertical motion. A turbulent process known as entrainment progressively assimilates the air in the free troposphere into the PBL.

The free atmosphere is where large-scale weather systems, such as jet streams, cyclones, and anticyclones, predominantly occur. It plays a crucial role in the movement and development of these weather patterns.

### 1.2.5 Entrainment Zone:

The capping inversion is not solid enough to trap all the thermals and turbulence. Sometimes the eddies overshoot creating a pressure gradient between the free atmosphere and the mixing layer. This causes an exchange of air, which traps some of the air parcels from the free atmosphere inside the mixing layer. This process is called entrainment, and the layer in which this takes place is called the entrainment zone.

Entrainment occurs whenever air from a nonturbulent region is drawn into an adjacent turbulent region. It is a one-way process that adds air mass to the turbulent mixed layer. It can be thought of as a mixed layer that gradually eats its way upward into the overlying air. It is responsible for the growth and evolution of the boundary layer from a mixed layer to a stable layer at night. It also plays a role in the development of clouds and precipitation, vertical distribution of temperature and humidity, dispersion of pollutants and aerosols, etc.

### 1.2.6 Residual Layer:

About half an hour before sunset the Earth cools down and the thermals stop forming. allowing turbulence to decay in the previously formed well-mixed layer. The resulting layer of air is sometimes called the residual layer because its initial mean state variables and concentration variables are the same as those of the recently decayed mixed layer. The residual layer is mutually stratified, resulting in turbulence that is nearly equal in intensity in all directions.

The residual layer is stable compared to the mixing layer and gradually decays into a more stable layer above the nocturnal boundary layer as time progresses to night.

### 1.2.7 Stable (Nocturnal) Layer:

As the night progresses, the bottom portion of the residual layer is transformed by its contact with the ground into a stable boundary layer. This is characterized by statically stable air with weaker, sporadic turbulence. The stable layer has little to no turbulence and no formation of thermals takes place.

The nocturnal boundary layer stays overnight and gradually decays into a mixed layer due to the entrainment of the free atmospheric air.

### 1.2.8 Significance:

Almost all of the weather and climate changes take place in the boundary layer and lower troposphere. BL plays a crucial role in transporting aerosols, clouds, and thermals throughout the atmosphere. Thus, the study of the boundary layer is very important and various research have been carried out regarding it. Wang et. al (2021) carried out an extensive study on BL and pollution using lidar and similar remote sensing methods. Wind tunnel studies have been used to observe the flow of neutral boundary layers over complex terrain and buildings, although the difficulty of stratifying the air has meant that typical daytime and nighttime boundary layers could not be adequately simulated (Stull, 1988).

To properly model the atmosphere, we need to understand the micrometeorology happening inside the boundary layer and ample refined data to support it.

## 1.3 CLOUDS

### 1.3.1 What is a cloud?

It's defined by the World Meteorological Organization (WMO) as a hydrometer consisting of a visible aggregate of minute particles of liquid water or ice, or both, suspended in the free air and usually not touching the Earth's surface. these atmospheric processes leading to cloud formation are important because there are several time scales on which these processes occur. In 1803 Luke Howard divided clouds into three categories:

1. Cirrus
2. Cumulus
3. Stratus
4. A fourth special type nimbus

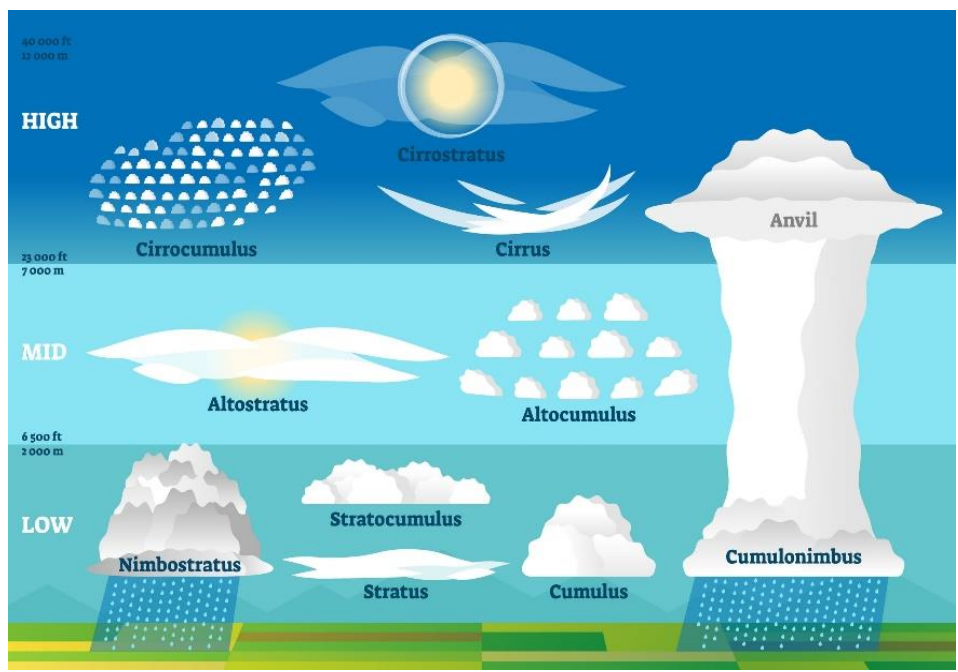

Figure 6. Different types of clouds observed. *Copyright:* normaals

## 1.3.2 Required elements for cloud formation

1. Cloud formation is moisture.

This moisture is constantly recycled through the earth-atmosphere system utilizing the hydrologic cycle Moisture in this cycle exists normally in the 3 states of water: solid, liquid, and vapor. The primary way to cool the atmosphere is through upward vertical motion or lifting of air.

2. Cloud formation is a source of lift in the following processes:

I. Fronts associated with low-pressure systems

II. orographic or mountain barriers

III. convection

IV. convergence (forced coming together of airflow)

This vertical motion or lift and the resulting cooling process. Similar to a hike in the mountains, it is cooler the higher one goes up in the atmosphere within the troposphere. The rate at which the air will cool with increasing altitude in the free atmosphere is referred to as the lapse rate or decrease of temperature with height.

The lapse rate of dry air is approximately 5.4°F/1000 feet or 9.8°C/1000 meters.

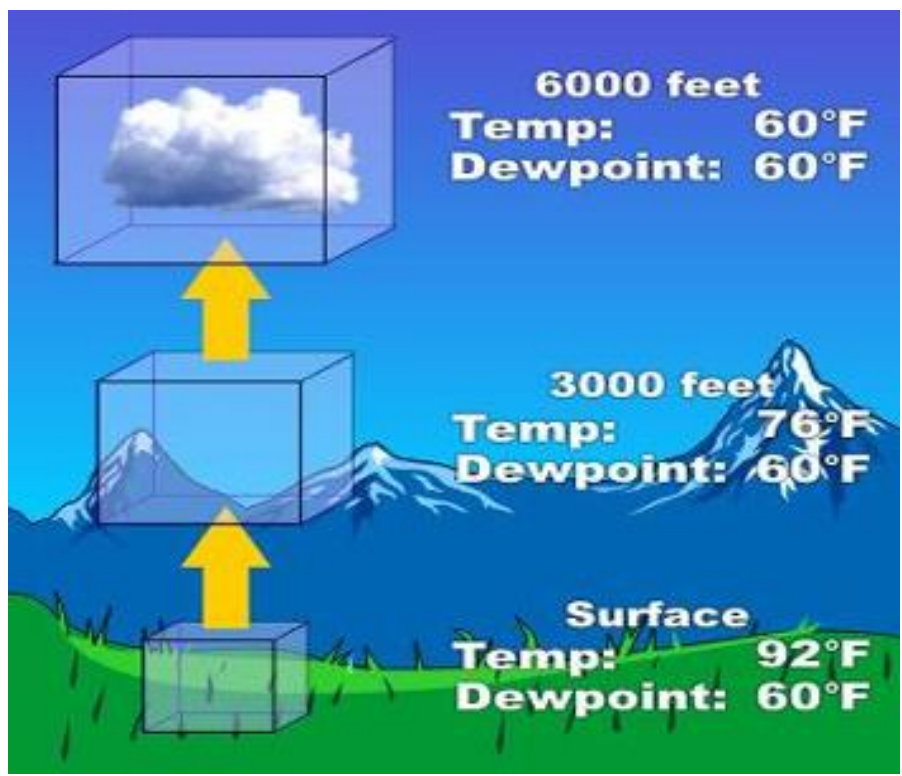

*Figure 7. Atmospheric cooling with increasing altitude (Source: Google Image)*

Mathematically gas laws form:

$$PV = RT \quad \text{or} \quad P = RT/V$$

( 4)

Where P = pressure

    V = molar volume

    T = temperature

    R = universal gas constant

when pressure decreases with increasing altitude and temperature decreases but the volume increases.

Radiational cooling is cooling in the lower layers of the atmosphere on clear, calm, and dry nights. The earth's surface cools, it will cool the air in contact with it. This air may be cooled to its saturation point resulting in the formation of late night or early morning fog or ground fog. This type of fog occurs frequently in river valleys

Main atmospheric processes creating atmospheric lift and clouds

### 1.3.3 Types of clouds

Cold air (dense) sinks and warm air (less dense) rises, clouds that form in an unstable environment (warm below and cold aloft) tend to be lumpy or globular in appearance. These clouds will resemble bubbles in a pot of boiling water. These are the cumuliform or convective clouds that we are all familiar with the localized nature of the up-drafts and down-drafts of convection.

**1.    Clouds due to lift by fronts**

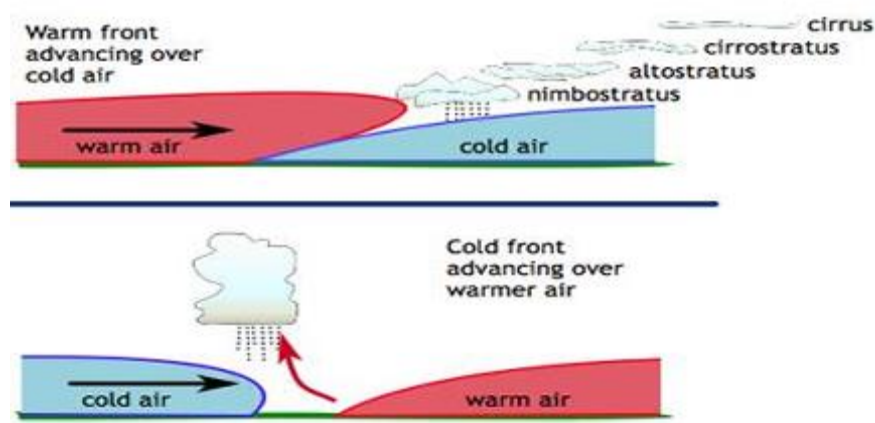

*Figure 8. Types of clouds due to the lifting of air in a region (Source: CMMAP)*

A warm front, both the warm advancing air and the cold retreating air are moving in the same direction. As warm air glides up and over cold surface air (warm front), the clouds tend to be layered or stratiform. As the front approaches, you observe a typical progression of clouds ranging from cirrostratus to altostratus, then further lowering and thickening to nimbostratus and steady precipitation. (as shown in the above figure) cold fronts cause more abrupt lifting with more intense localized vertical motion as the cold and warm air masses collide. This generally results in cumuliform clouds with showery conditions as the cold air undercuts and forces the warm air up.

## 2. Orographic lift clouds

Airflow perpendicular to a range of hills or mountains is forced to rise up and over the mountains. the air rises on the windward side of the mountain range it cools and may eventually reach its saturation point with clouds forming. The reverse is true as the air descends down the leeward side of the mountains. Then the air is compressed. warming air can hold more moisture before reaching saturation. As a result, clouds tend to break up into the lee of mountains.

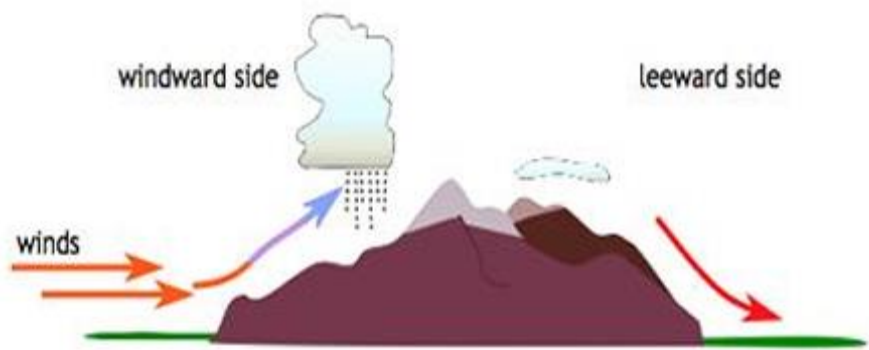

*Figure 9. Orographic lift of clouds due to the presence of mountains (Source: CMMAP)*

This process during a winter snowstorm, with heavy snow along the windward side and lesser amounts to the leeward side of the mountains.

## 3. Lift due to convection

The white cotton ball (cumulus) type clouds on a warm summer afternoon. This is the process of convection.

The Earth's atmosphere is transparent to incoming solar radiation. This radiation hits the ground, it's converted to heat energy. As the ground warms, the air in contact with the ground is also warmed through conduction.

The air is warmed, it becomes less dense, thus it rises convection. Air rises it cools, with clouds ultimately forming over the updraft as depicted. The spacing of these up and down drafts results in the observed distribution of cumulus clouds. On the edges of the clouds, cool air sinks to replace the warm air rising, thereby completing the convection cell.

4.     **Convergence and lift**

Another source of lift, which is a combination of the above processes, is convergence. When air is forced to converge or come together (such as at the center of a low-pressure system), it can only go upward (can't go into the ground). An example would be the air flowing inward toward the center of low pressure which is forced to rise.

### 1.3.4   Measurement of the effect of smoke on inhibition of cloud formation:

Air pollution and smoke from fires have been modeled to reduce cloud formation by absorbing sunlight, thereby cooling the surface and heating the atmosphere. Satellite data over the Amazon region during the biomass burning season showed that scattered cumulus cloud cover was reduced from 38% in clean conditions to 0% for heavy smoke. This response to the smoke radiative effect reverses the regional smoke instantaneous forcing of climate from–28 watts per square meter in cloud-free conditions to 8 watts per square meter once the reduction of cloud cover is accounted for.

Aerosols on the atmospheric radiation budget and climate constitute the greatest uncertainty in attempts to model and predict climate. (CHANGE, 2007) Aerosols can counteract regional greenhouse warming by reflecting solar radiation to space or by enhancing cloud reflectance or lifetime. (Twomey, 1977) However, aerosol absorption of sunlight is hypothesized to slow down the hydrological cycle and influence climate in ways not matched by the greenhouse effects. (Ramanathan, 2001) During periods of heavy aerosol concentration over the Indian Ocean and Amazon basin. (Reid, 1998)

Less irradiation of the surface means less evaporation from vegetation and water bodies, a more stable and drier atmosphere, and consequently less cloud formation. This effect was defined theoretically as positive feedback to aerosol absorption of sunlight and was termed the semi-direct effect. (Hansen, 1997) A similar process, defined as cloud burning by soot, in which solar heating by the aerosol reaches its maximum near the top of the boundary layer, thereby stabilizing the boundary layer and suppressing convection, has been described. (Ackerman B., 2000)

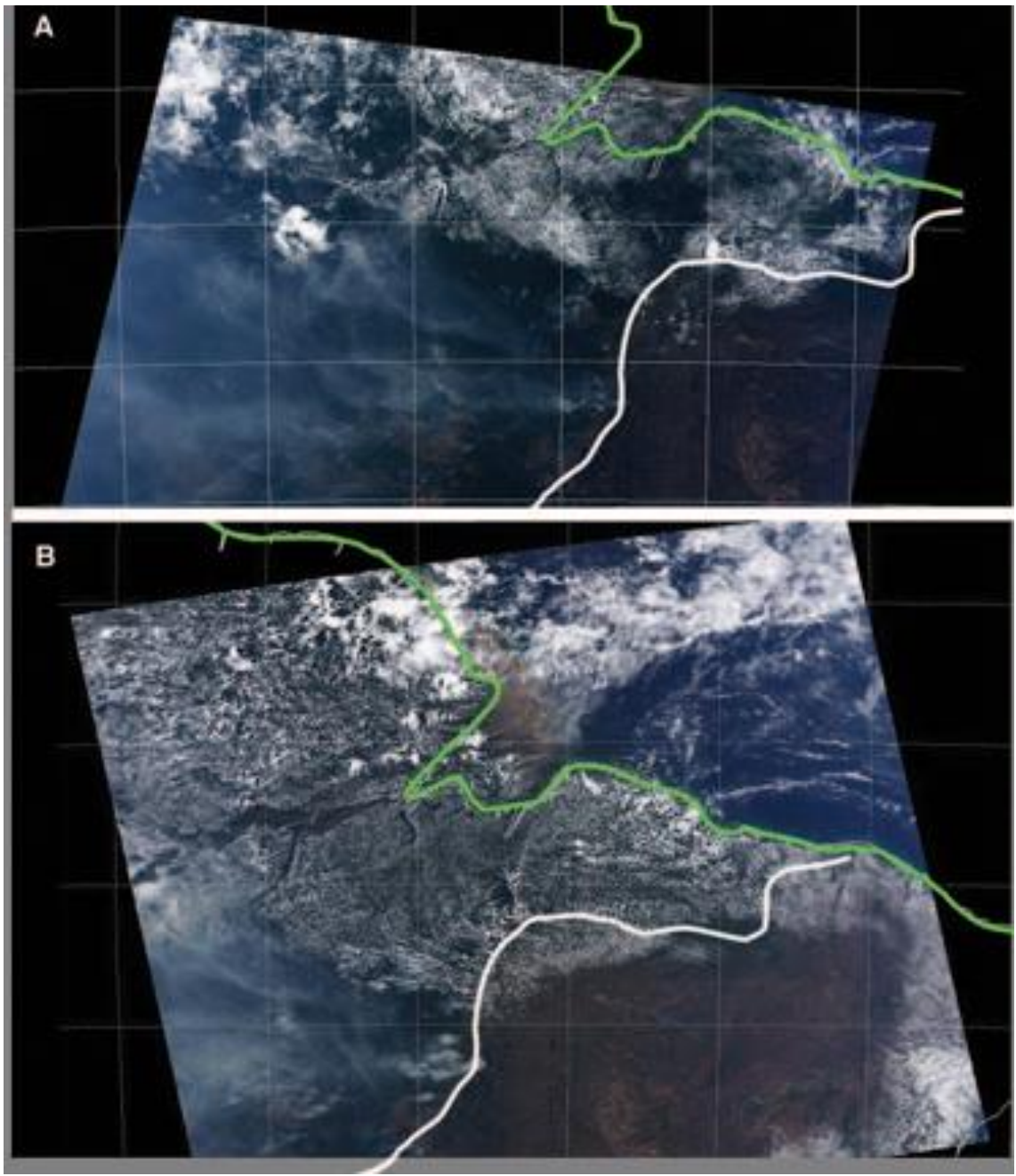

*Figure 10. Terra and Aqua satellite images of the east Amazon basin, 11 August 2002. The boundary between the forest and the Cerrado region is marked in white on both images, and the seashore is marked in green. Note that the Amazon River and its tributaries are a cloud-free area. Koren et al. (2004).*

Cloud simulations were based on aerosol observations of INDOEX (Indian Ocean Experiment) (Heymsfield, 2001) and focused mainly on the amplification of daytime clearing due to aerosol heating. The reduction of evaporation from the Mediterranean Sea by pollution from northern and eastern Europe was modeled to reduce cloud formation.

The scattered cumulus clouds emerge regularly in the morning over the eastern shore. By local noon they cover large parts of the Amazon basin. The cloud diameter is on average 2 to 3 km, with average reflectance of 0.35 in the visible part of the solar spectrum. The clouds are sensitive to the surface properties and seldom form directly above the Amazon River. The stable meteorological conditions and the regular behavior of the clouds in the absence of smoke create an ideal case to study the impact of aerosol from biomass burning on cloud formation. Satellite images of the absence of clouds in smoky regions and differences between morning and afternoon are shown in the below figure. In the afternoon, clouds uniformly cover the region of the Amazon ozone forest that is not filled with high con concentrations of smoke (Ackerman A., 2000).

The estimate the effect of smoke on cloud coverage and the semi-direct forcing on climate we measured the cloud fraction as a function of the smoke optical depth in the cloud vicinity.

The isolated effect of smoke on cloud fraction it is necessary to minimize any residual influences of synoptic condition and variation of surface cover.

Clouds are detected and separated from smoke by their high local spatial variability and brightness reflectance. (Martins, 2002) Size and spatial variability were then used to classify between the small scattered cumulus and other types of clouds. The smoke ODs were calculated by the MODIS algorithm in the noncloudy area. (Kaufman, 1997)

## 1.4 REMOTE SENSING:

Remote sensing is the process of acquiring information from a source without being in contact with it. This is done by sensing and recording the reflected or emitted energy and processing, analyzing, and applying the information. The process involves interaction between the incident radiation and the target of interest.

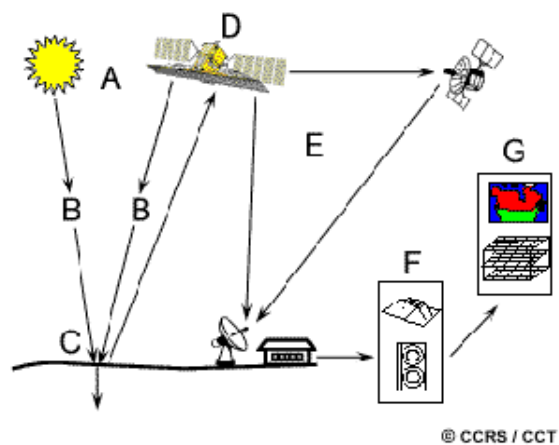

*Figure 11. The basic elements required for the remote sensing process. (Source: Canada Centre for Remote Sensing)*

The basic elements of a remote sensing process from beginning to end are (Figure 11):

    (A) Energy source or illumination
    (B) Radiation and the atmosphere
    (C) Interaction with the target
    (D) Recording energy by the sensor
    (E) Transmission, reception, and processing
    (F) Interpretation and analysis of the data

(G) Application

Remote sensing is mainly used for getting information from places where in-situ measurements are very hard to carry out. Most of the remote sensing is done by electromagnetic waves that are reflected and absorbed to construct a digital image of the target. Some EM waves are absorbed or reflected by the atmospheric components, like water vapor and carbon dioxide, while some wavelengths allow for unrestricted movement through the atmosphere; visible light has wavelengths that can be transmitted through the atmosphere. The microwave spectrum has wavelengths that can pass through clouds, an attribute utilized by many weather and communication satellites.

### 1.4.1 Types of Remote Sensing:

According to the source for remote sensing, it can be divided into two categories i.e. Active Remote Sensing and Passive Remote Sensing.

Active remote sensing has sensors that can emit their own electromagnetic radiation on the target and observe the reflected light from it. Whereas, the in passive remote sensing the source of energy is not the sensor itself. Mainly the sun is used for the source of passive remote sensing.

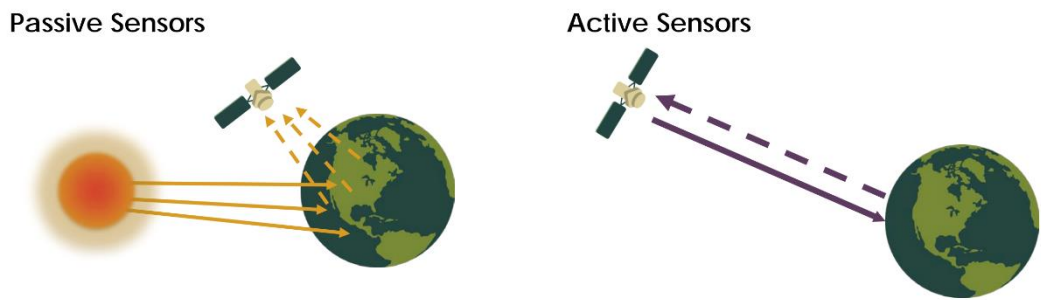

*Figure 12. Active and Passive Remote Sensing (Source: NASA Science).*

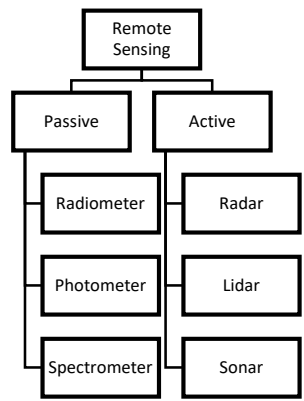

*Figure 13. Examples of the Active and Passive Remote Sensing*

The remote sensing equipment can be placed in ground, sky, and space. Each have their benefits and shortcomings. Ground-based sensors are often used to record detailed information about the surface which is compared with information collected from aircraft or satellite sensors. In some cases, this can be used to better characterize the target which is being imaged by these other sensors, making it possible to better understand the information in the imagery.

Aerial platforms are primarily stable-wing aircraft, although helicopters are occasionally used. Aircraft are often used to collect very detailed images and facilitate the collection of data over virtually any portion of the Earth's surface at any time. UAVs and other RC drones are actively used to carry LIDAR (LIght Detection And Ranging), infrared sensors, etc. for military purposes. Integration of machine learning with these drones are being used to detect and track moving targets (Li, Zhao, Zhang, & Tan, 2018).

In space, remote sensing is sometimes conducted from the space shuttle or, more commonly, from satellites. Satellites are objects which revolve around another object - in this case, the Earth. For example, the moon is a natural satellite, whereas man-made satellites include those platforms launched for remote sensing, communication, and telemetry (location and navigation) purposes. Because of their orbits, satellites permit repetitive coverage of the Earth's surface on a continuing basis. Cost is often a significant factor in choosing among the various platform options.

### 1.4.2 Radar and Lidar:

Radar (Radio detection and ranging) is a type of active remote sensing which uses electromagnetic energy backscattered from ground targets to extract physical and di-electrical properties of the specified target. Modern radars use electromagnetic waves in the microwave spectrum for collection of data. Pulses of radiation are emitted with a fixed timespan, which is backscattered and reflected back to the sensor by the target to be measured. Because RADAR provides its own energy source, images can be acquired day or night. Also, microwave energy is able to penetrate through clouds and most rain, making it an all-weather sensor.

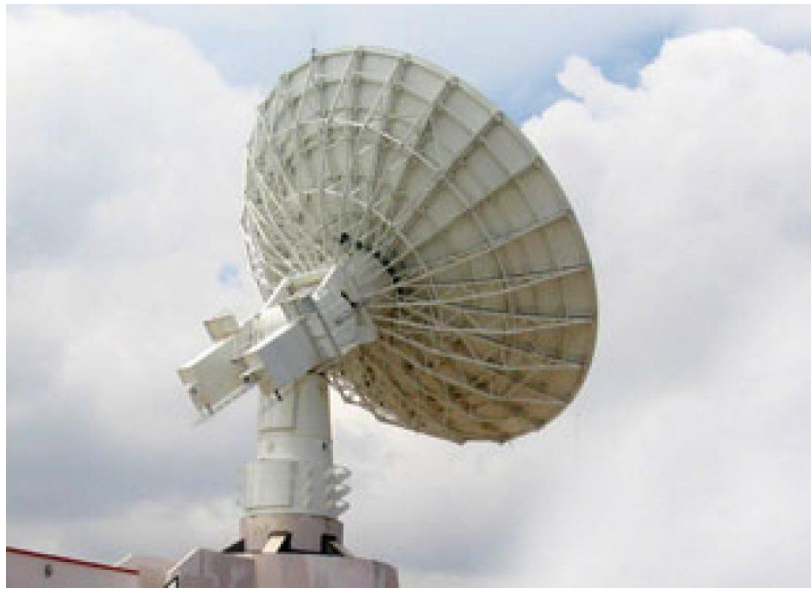

*Figure 14. The Indian Space Research Organization on May 28 inaugurated a new navigation hub in the Indian Deep Space Network (IDSN) complex at Byalalu, about 40 km from Bangalore.*

Lidar stands for Light Detection And Ranging. It uses a laser beam to measure the distance and shape of objects in the environment.

A lidar determines the distance of an object or a surface with the formula:

$$d = c \frac{\Delta t}{2}$$

*( 5)*

where c is the speed of light, d is the distance between the detector and the object or surface being detected, and Δt is the time spent for the laser light to travel to the object or surface being detected, and then travel back to the detector.

Lidar has a very high accuracy and resolution but is affected by atmospheric conditions like fog and rain. It is mainly used in Topographical mapping, autonomous vehicles, archaeology, and weather studies.

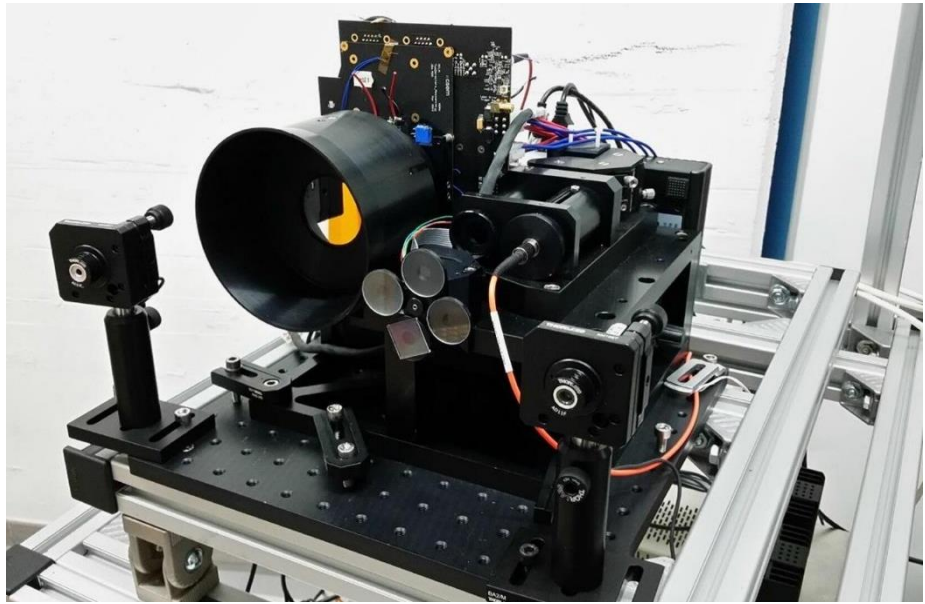

*Figure 15. Lidar system manufactured by the ESA (European Space Agency) in 2018.*

### 1.4.3  Various remote sensing satellites in use:

**GOES:** The GOES (Geostationary Operational Environmental Satellite) System is the follow-up to the ATS series. They were designed by NASA for the National Oceanic and Atmospheric Administration (NOAA) to provide the United States National Weather Service with frequent, small-scale imaging of the Earth's surface and cloud cover.

**NOAA AVHRR:** NOAA is also responsible for another series of satellites that are useful for meteorological, as well as other, applications. These satellites, in sun-synchronous, near-polar orbits (830- 870 km above the Earth), are part of the Advanced TIROS series (originally dating back to 1960) and provide complementary information to the geostationary meteorological satellites (such as GOES).

**IRS:** The Indian Remote Sensing (IRS) satellite series, combines features from both the Landsat MSS/TM sensors and the SPOT HRV sensor. The third satellite in the series, IRS-1C, launched in December 1995 has three sensors: a single-channel panchromatic (PAN) high-resolution camera, a medium resolution four-channel Linear Imaging Self-scanning Sensor (LISS-III), and a coarse resolution two-channel Wide Field Sensor (WiFS).

**INSAT:** The Indian National Satellite System or INSAT, is a series of multipurpose geostationary satellites launched by the Indian Space Research Organization (ISRO) to satisfy telecommunications, broadcasting, meteorology, and search and rescue operations.

**AstroSat:** AstroSat is India's first dedicated multi-wavelength space telescope. It was launched on a PSLV-XL on 28 September 2015. With the success of this satellite, ISRO has proposed launching AstroSat-2 as a successor for AstroSat. AstroSat performs multi-wavelength observations covering spectral bands from radio, optical, IR, UV, and X-ray wavelengths.

**SABER:** SABER (Sounding of the Atmosphere using Broadband Emission Radiometry) is a 10-channel infrared radiometer satellite to study the structure, energetics, chemistry, and dynamics of the Earth's mesosphere and lower thermosphere. SABER provides a never-before-seen view of the atmosphere and paves the way for a new area of science. SABER, built by Utah State University Space Dynamics Laboratory and managed by NASA Langley Research Center, is one of four instruments on the TIMED (Thermosphere, Ionosphere, Mesosphere, Energetics, and Dynamics) spacecraft, which launched in late 2001.

**MODIS:** MODIS (or Moderate Resolution Imaging Spectroradiometer) is a key instrument aboard the Terra (originally known as EOS AM-1) and Aqua (originally known as EOS PM-1) satellites. Terra MODIS and Aqua MODIS are viewing the entire Earth's surface every 1 to 2 days, acquiring data in 36 spectral bands, or groups of wavelengths (see MODIS Technical Specifications). These data will improve our understanding of global dynamics and processes occurring on the land, in the oceans, and in the lower atmosphere.

# 2 CHAPTER 2

## 2.1 LITERATURE REVIEW:

### 2.1.1 Studying Trends and Changes in The Atmosphere:

Air motions play a major role in determining the distributions of chemical components in the atmosphere. These motions are determined mainly by three principal forces: gravity, pressure-gradient, and Coriolis (Jacob, 1999). The dynamics of the atmosphere keep on changing in a timescale of less than an hour. Thus, to understand the atmosphere, especially its present state and future evolution, we need to observe and measure its different elements both at the surface and upper levels. However, there is a difficulty here. Unlike air in a closed laboratory, the free atmosphere is constantly in a turbulent motion. So, what is observed or measured at a particular time is not representative of what comes over after an interval of time (Saha, 2008).

For an accurate description of the atmosphere, we need to solve numerous equations of motion of aerosols and other particles, consider all the atmospheric forcings, and take into account various instantaneous natural as well as anthropogenic changes. It is computationally impossible to create a perfect model of the Earth's atmosphere. H. Hersbach (2020) and Bell et al. (2021) analysed the effectiveness of ERA5 (The upgraded version of ERA-Interim) in observing atmosphere, land, and ocean waves. The addition of observed data from various satellites and the assimilation of these datasets enhances the predictability. However, larger cold bias in the lower stratosphere, unnecessarily high precipitation called 'rain bombs', unrealistic snow depth, etc. are shortcomings of the model. Hence, observed data is required to properly calibrate the models.

Esplin et al. (2023) studied the lower atmosphere using SABER and argued it to be stable in its calibration after more than 21 years in orbit despite having an originally planned mission life of 2 years. Mlynczack et al. (2020) argued that the single most important decision made in the development of SABER was to commit to producing the most accurately calibrated instrument possible for the available resources. This one decision guided parts development, parts testing and selection, instrument thermal and mechanical design, and instrument operations. The decision to focus on calibration resulted in an instrument that is remarkably stable as discussed in detail in Mlynczack et al. (2020).

Ichoku et al. (2002) analysed the characteristics of the 5-channel MICROTOPS II sun photometer for measuring AOT and water vapour. They studied 5 sun photometers and compared the observed data to a better version of the AERONET Sun photometer to calibrate them. Excluding the error due to operation and cleanliness of the lens Microtops can be quite accurate and stable, with root-mean-square (rms) differences between corresponding retrievals from clean calibrated Microtops and the AERONET Sun photometer being about ±0.02 at 340 nm, decreasing down to about ±0.01 at 870 nm. Gong et al. (2018) confirmed the bias to be in the range of -0.043 to -0.005 which could be due to improper pointing at the sun or unstable electronics. Thus, measurements done by it have good precision and accuracy with relatively less uncertainty.

Münkel et al. (2007) studied the boundary layers and dust concentration using Vaisala ceilometers and found its enhanced optics and electronics enable the CL31 ceilometer to detect fine boundary-layer structures whose counterparts are seen in temperature profiles. Gelaro et al. (2017) and Hutchison et al. (2003) used MERRA-2 and MODIS data respectively to study various trends in the climate like aerosol and air quality data and have explained the significance of both reanalysis and remote sensing data for atmospheric studies.

### 2.1.2 Dust Storm Over Ahmedabad (May, 2024):

According to the Earth Observatory website http://earthobservatory.nasa.gov/ dust storms are considered natural hazards, which affect ecosystems for a short time interval ranging from a few hours to a few days. Due to the adverse effects of the dust storms on climate and human health, Kaskaoutis et al. (2010) have observed a 3 times increase in the research about these events in the last two decades. Engelstaedter et al. (2006) state that these studies shed light on the dust source regions, the variability of dust emissions, dust transport, the role of human impact and rainfall on dust emission, and the recent developments of global and regional dust models.

Bian et al. (2011) used a multi-step approach to study a severe dust storm (SDS) event that took place in China in the year 2010, integrating data from MODIS and the 'WRF-Dust' dust model to analyse the transport of dust over China. The calculation showed that the calculated dust concentrations were considerably lower than the measured values in the downwind regions of deserts when the propagate dust source (PDS) process was not included in the model. PDS is a process where a high concentration of dust particles in a region acts as a secondary source of the dust storm. This process was previously hinted at by Engelstaedter et al. (2006) using the TOMS data over North African regions.

Baddock et al. (2014) studied the correlation between dust concentration and visibility in the range of 10-100 km over Australia. They found a strong positive correlation where visibility is severely affected by the distance from the source, concentration, and size of the dust particles. Kotthaus et al. (2018) observed the boundary layers of central London using lidar and ceilometer for pollution dispersion and structure of boundary layer in polluted cities. They found that the presence of turbulence in the boundary layer causes an increase in aerosol in the atmosphere in highly polluted cities.

Littmann (1990) argued that the results from the cross-correlation of mean values show that there is no universal pattern of interrelation but regional patterns correspond with regional climatic dynamics. Apart from some exceptions, there is only a weak dependency of storm frequencies on mean precipitation, temperature, wind, and evaporation data.

Singh et al. (2022) measured the particulate matter $PM_{2.5}$ and AOD of a severe dust storm over the central Himalayas during June 13–17, 2018. They observed a sudden hike in the concentration during the SDS event.

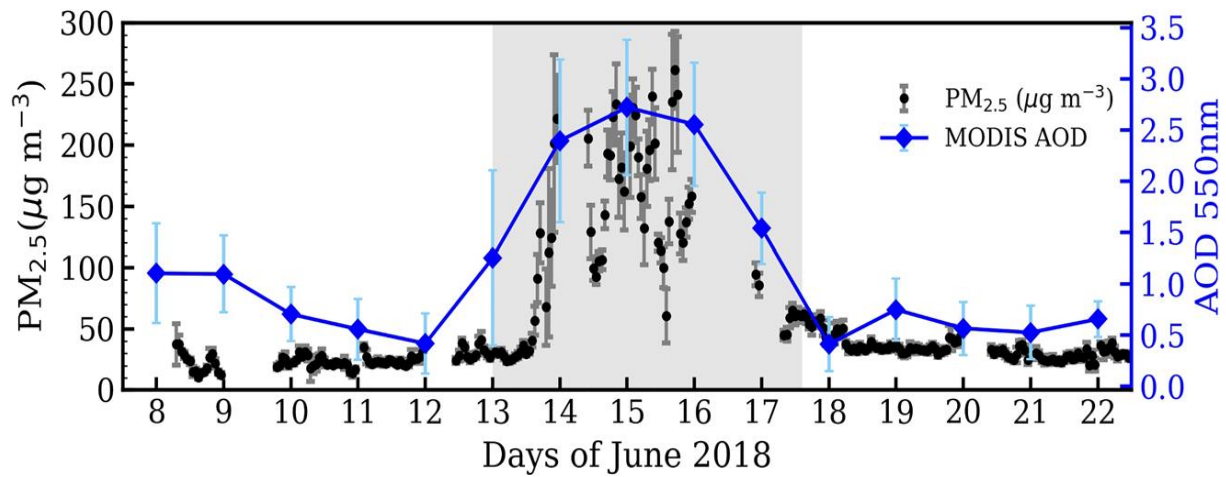

Figure 16 Variations in hourly PM2.5 and Moderate Resolution Imaging Spectroradiometer (MODIS) retrieved daily Aerosol Optical Depth (AOD)550 over Manora Peak in the central Himalayas. The shaded region marks the peak dust impact period (Singh J. S., 2022).

Chakravarty et al. (2021) studied dust storms in Northern India and concluded the source of these storms to be the Thar Desert. Dust storms and their transport to the Northern Indian Plains (or Indo-Gangetic Plains; IGP) are frequently witnessed during the pre-monsoon months (March through May). They have used Doppler Weather Radar (DWR) and MODIS together with ERA5 to observe aerosol and cloud characteristics with high accuracy. The large-scale circulation before the dust storm event showed a mixing of low-level warm air from the Middle East and dry cold air which resulted in a potentially unstable convective environment and an ideal background for triggering thunderstorms over the region. The increase in particulate matter $PM_{10}$ and $PM_{2.5}$ was observed to magnitudes as high as 950 $\mu g m^{-3}$ and 550 $\mu g m^{-3}$ respectively, just at the incipient stage of the dust storm.

Recently, Saha et al. (2022) observed a dust storm over Ahmedabad (23.02°N, 72.57°E) on 27 April 2021 using an all-weather Ceilometer Lidar (CL31) operational at Physical Research

Laboratory (PRL), Ahmedabad as well as taking data from MODIS, ERA5, and Coupled Ocean-Atmosphere Radiative Transfer (COART) model. They observed a blanket of dust covering the region from MODIS data.

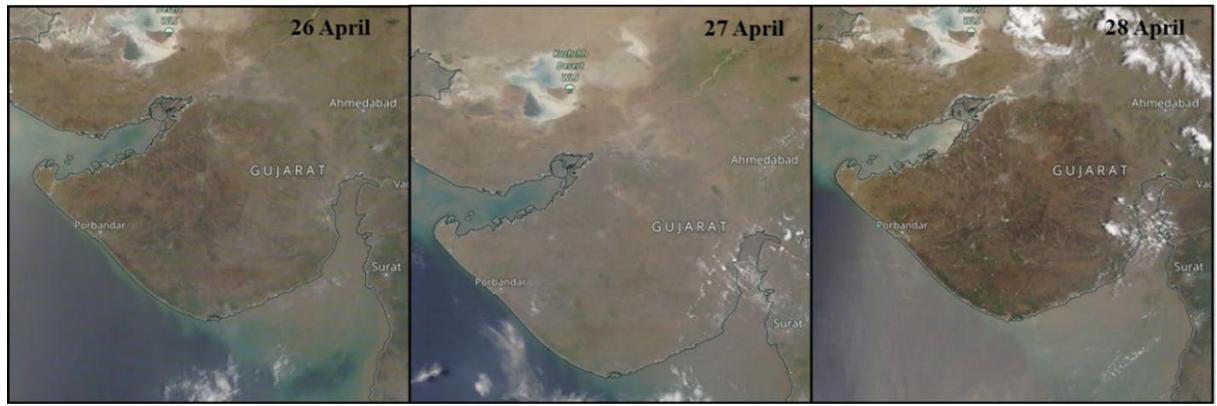

*Figure 17. Visual satellite images obtained from MODIS onboard TERRA on 26–28 April 2021. The image obtained on 27 April shows a blanket of dust covering the observation site (Saha, et al., 2022).*

A drop in the surface temperature was reported due to a reduction in solar radiation by the dust coverage. Humidity as well as the concentration of particulate matter increased drastically during the dust storm event studied by them. The collapse of the nocturnal boundary layer was seen due to rapid vertical mixing. The temperature drop was verified by the observations previously reported by Kedia et al. (2018) who used the WRF-CHEM model for the SDS event in India.

**2.1.3   Volcanic Eruption in Mount Ruang, Indonesia (April, 2024):**

Volcanoes are a major source of sulphur, acids, aerosols, and other gases that have an atmospheric impact. Von Glasow et al. (2009) studied the chemicals released from a volcano, their mechanism, and their effects on the atmosphere. The adverse effects on the climate are caused by sulphate aerosols found in troposphere. They scatter the sunlight back to space cooling the Earth's surface. However, Mass et al. (1989) did an epoch analysis of the major volcanic eruptions, where he had found that post-eruption cooling is unwarranted. Another effect of this sulphate is to contribute to cloud condensation nuclei (CCN), which facilitates precipitation. Von Glasow et al. (2009) also studied the acid deposition in the atmosphere due to volcanic events. Arndt et al. (1997) estimated, based on data for 1987–1988, that deposition of SO2 from volcanoes accounts for about 30% of the total $SO_2$ deposition in Japan, 50% in Indonesia, and 20% in the Philippines, respectively. Baxter et al. (2015) analysed the effects of volcanic eruptions on human health and listed out the main factors of causalities i.e. pyroclastic density currents, lahars, explosions, tephra fall, gases, and gas-ash interaction.

Pergola et al. (2001) used NOAA/AVHRR data to propose a robust technique to monitor volcanic aerosols. Khokhar et al. (2005) observed $SO_2$ from 1996 to 2002 using Global Ozone Monitoring Experiment (GOME) data to show the effectiveness of remote sensing equipment in studying such events. Vernier et al. (2011) examined the effect of eruptions on stratospheric aerosol using SAGE II, CALIPSO, and GOMOS in the last decade. They observed as much as two times an increase in the concentration of aerosols and a hike in the $SO_2$ entering the stratosphere.

Hariyono et al. (2018) studied the characteristics of the volcanoes in Indonesia. They classified 147 volcanoes, 76 of which are active volcanoes and spread along the islands of Java, Lesser Sunda, Sumatra, and Celebes. Most of these are stratovolcanoes including Mount Ruang but do not always show explosive eruptions. Kushendratno et al. (2012) conducted a case study of Soputan Volcano in Indonesia using satellite data as well as field observation. They had similar findings, such as an increase in aerosol and $SO_2$ concentration, as well as earthquakes before and after the eruptions.

# 3 CHAPTER 3

## 3.1 METHODOLOGY:

### 3.1.1 Instruments and their description:

#### 3.1.1.1 Sun photometer

The instrument is equipped with five (5) accurately aligned optical collimators, capable of a full field view of 2.5°. Internal baffles are also integrated into the device to eliminate internal reflections. Each channel is fitted with a narrow-band interference filter and a photodiode suitable for the particular wavelength range. The collimators are encapsulated in a cast aluminum optical block for stability.

A sun target and pointing assembly is permanently attached to the optical block and laser-aligned to ensure accurate alignment with the optical channels. When the image of the sun is centered in the bull's-eye of the sun target, all optical channels are oriented directly at the solar disk. A small amount of circumsolar radiation is also captured, but it makes little contribution to the signal. Radiation captured by the collimator and bandpass filters radiate onto the photodiodes, producing an electrical current that is proportional to the radiant power intercepted by the photodiodes.

These signals are first amplified and then converted to a digital signal by a high-resolution A/D converter. The signals from the photodiodes are processed in series. However, with 20 conversions per second, the results can be treated as if the photodiodes were read simultaneously. The ozone layer, a concentration of 3-atom oxygen molecules in the stratosphere, is essential to life on Earth. Short wavelengths of ultraviolet radiation are much more readily absorbed by ozone than the longer wavelengths in the same UV bandwidth. This means that the amount of ozone between the observer and the Sun is proportional to the ratio of two wavelengths of the Sun's ultraviolet radiation.

MICROTOPS II uses that relationship to derive the Total Ozone Column (the equivalent thickness of pure ozone layer at standard pressure and temperature) from measurements of 3 wavelengths in the UV region. Similarly, as in the traditional Dobson instrument, the measurement at an additional 3rd wavelength enables a correction for particulate scattering and stray light.

The precipitable water column is determined based on measurements at 936nm (water absorption peak) and 1020nm (no absorption by water). The aerosol optical thickness at 1020nm is calculated based on the extraterrestrial radiation at that wavelength, corrected for the sun-earth distance, and the ground-level measurement of the radiation at 1020nm.

This instrument is designed to allow quick and inexpensive measurements of AOT (aerosol optical thickness). This instrument stores 800 scans of all wavelengths along with date, and time and there is a GPS connected that measures the latitude and longitude of the given place. It also calculates the solar zenith angle (SZA). The range of SZA is from 0-90 degrees.

When a sun photometer is placed somewhere within the earth's atmosphere, the measured radiance is not equal to the radiance emitted by the sun because the solar flux is reduced by atmospheric absorption and scattering. When the image of the sun is placed in the bull's eye of the sun target the radiation captured by the collimator and the band pass filter radiates onto the photodiode causing a photocurrent proportional to radiant power. These signals are amplified and converted into digital signals and then stored in memory.

The popularity of Microtops sun photometers is due to their ease of use, portability, and relatively low cost. The instrument has five wavelengths that can be chosen based on the interference filter installed. Depending on the filter combination the instrument is designed to measure aerosol optical depths column ozone concentration and column water vapor concentrations. The system uses a photodiode detector coupled with a full field of view of 2.5°.

It must be connected to a PC running the operating software in order to make measurements and store data. The software allows comprehensive user selection of the scanning modes. It uses a filter wheel and photodiode detector for measurements for sun tracking tripod base, sun sensor, and rain sensor are included.

The equation $I = I_0 e^{-at}$ describes theoretically how to interpret sun photometer measurements requires that the instruments should see only direct light from the sun that is light that follows a straight-line path from the sun to the light detector. Sun photometers will see some scattered light and direct light from the sky around the sun. the cone of light a sun photometer detector is called its field of view and it is desirable to have this cone as narrow as possible. The sun photometer's field of view is about 2.5 degrees and concluded to be a reasonable compromise between desires for accuracy and practical considerations in building a handheld instrument. The basic tradeoff is that the smaller the field of view, the harder the instruments are to point accurately at the sun. very expensive sun photometers with motors and electronics to align the detector with the sun typically have a field of view of 1 degree or less.

Sun photometer is used to measure aerosol optical depths to measure the column aerosol optical loading. Sun photometer consist of a narrow field of view sensor that is pointed at the sun. many sun photometers are handheld, the more expensive systems are automated and point to the sun automatically.

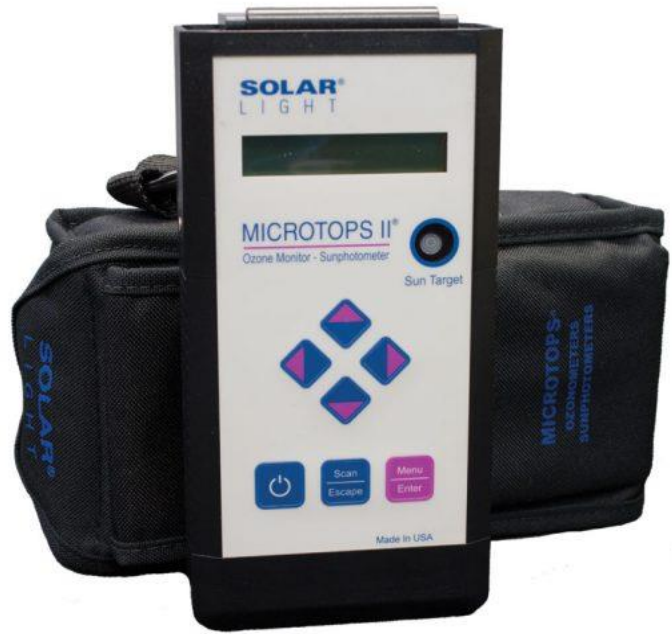

*Figure 18.MICROTOPS II Ozone Monitor & Sunphotometer, Version 2.43 Revision C (Source: Microtops user manual)*

To measure the optical, physical, and radiative properties of atmospheric aerosols, the worldwide AErosol Robotic NETwork, AERONET, has been established, with more than 1000 automatic sun-photometer (SPM) stations. (Holben, 1998)

The most famous SPMs is Calito, which measures the sun's irradiance in optional wavelengths and directly calculates the AOT. Therefore, it can be determined the rate of aerosols in the atmosphere and characterizes their size distribution such as smoke, polluting gases, ice crystals, and dust (Sharma, 2014). The weather by scattering or absorbing solar and thermal radiation and changing cloud characteristics in different ways. (Lohmann, 2005)

Aerosol optical thickness, AOT, indicates the amount of sunlight absorbed or scattered by atmospheric particles. It is an essential parameter for visibility degradation (due to atmospheric particles), extinction of solar radiation, climate effects, and tropospheric corrections in remote sensing. (Dubovik, 2002)

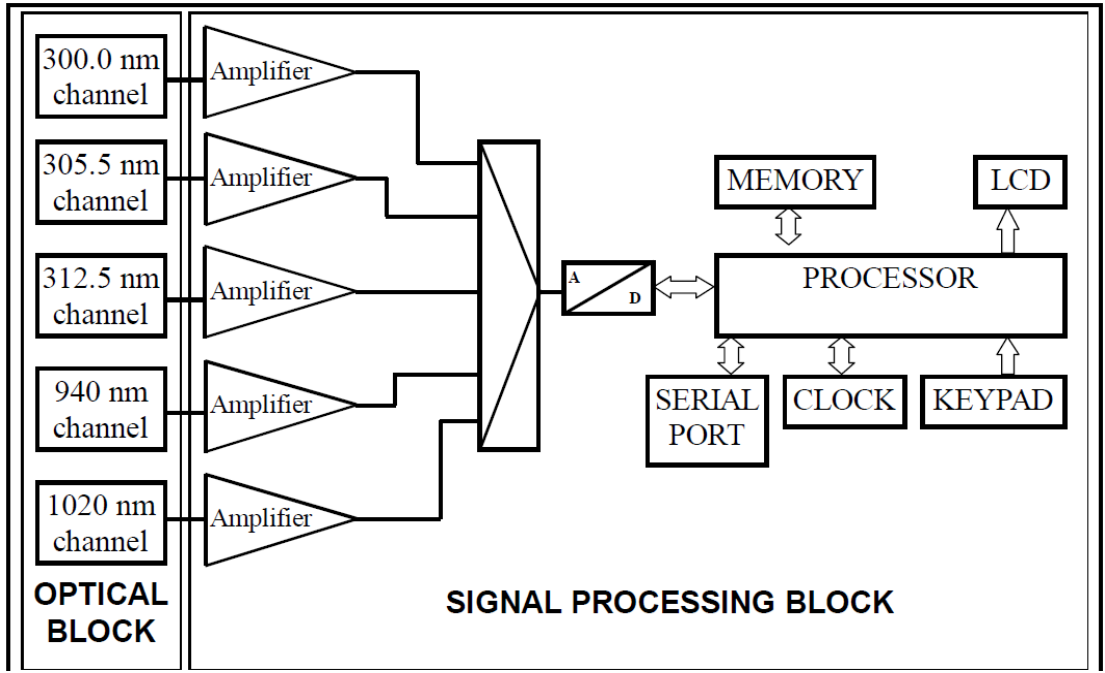

*Figure 19. MICROTOPS II Structure (Source: Solar Light Company, Inc.)*

**Calculation of the aerosol optical thickness**

Microtops handheld SPM records the sunlight voltage, $V\lambda$, at 5 channels with 300, 305.5, 312.5, 940, and 1020 nm wavelength channels. These wavelengths have dynamic molecular absorption. Aerosol optical thickness (AOT) values are calculated using the following equation:

$$\tau_a(\lambda) = \frac{[\ln(V_0(\lambda) \times (\frac{r_0}{r})^2) - \ln(V(\lambda))]}{m} - a_R(\lambda) \times \frac{P}{P_0} - \tau_{O_3}(\lambda),$$

*( 6)*

where $V0(\lambda)$ is the calibration constant,

$r_0$ is the distance of 1 Astronomical Unit (AU),

r is the Sun-Earth distance at the recording date (in AU).

$m$ is the optical air mass,

P is the pressure at the observation point,

$P_0$ is the pressure at the sea level,

$a_R(\lambda)$ indicates the extinction of sunlight due to Rayleigh scattering,

In our study, readings were taken from the MICROTOPS II sunphotometer every day from 15[th] June at 2-hour intervals. Which is as follows: 10:00 am, 12:00 pm, 1:00 pm and 4:00 pm. We took three readings at the same time to minimize the error due to changing atmospheric parameters. Then the average of the three consecutive data was used. The sunphotometer was pointed directly at the sun and the light was calibrated through a small hole that prohibits scattered sunlight from reaching the instrument's detector.

The data collected was used to create a time series plot of the parameters Ozone concentration, Water column thickness, and Aerosol optical thickness to compare with the satellite data.

### 3.1.1.2 Ceilometer

A ceilometer is a device for measuring the height of cloud bases and overall cloud thickness. The device works as day or night by shining an intense beam of light a light produced by an infrared or ultraviolet transmitter or a laser, modulated at an audio frequency, at overhead clouds. Reflections of this light from the base of the clouds are detected by a photocell in the receiver of the ceilometer. There are two basic types of ceilometers: the scanning receiver and the rotating transmitter.

The scanning-receiver ceilometer has its separate light transmitter fixed to direct its beam vertically. The receiver is stationed a known distance away. The parabolic collector of the receiver continuously scans up and down the vertical beam, searching for the point where the light intersects a cloud base. When a reflection is detected, the ceilometer measures the vertical angle to the spot; a simple trigonometric calculation then the height of the cloud ceiling. Many modern scanning-receiver ceilometers use a laser pulse to identify the height of a cloud's base and top and various points in between to create a vertical profile of the cloud. The rotating transmitter ceilometer has its separate receiver fixed to direct reflections only from directly overhead while the transmitter sweeps the sky. When the modulated beam intersects a cloud base directly over the receiver, light is reflected downward and detected.

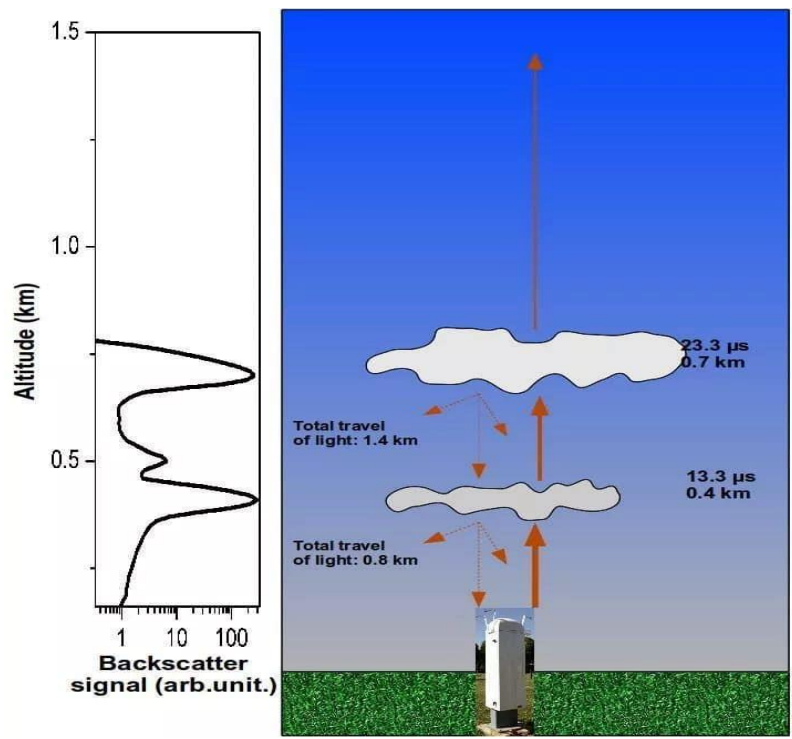

*Figure 20. Working of Ceilometer and backscatter profile. (Source: www.askpilot.info/2020/06/what-is-ceilometers.html)*

The planetary boundary layer (PBL) is the layer where the earth's surface interacts with the large-scale atmospheric flow. Since substances emitted into this layer disperse gradually horizontally and vertically through the action of turbulence and become completely mixed if sufficient time is given and sinks and sources are absent, this layer is also called the mixing layer.

The PBL height or mixing height (MH) is a key parameter in air pollution models determining the volume available for pollutants to dispersion (Seibert, 2000) and the structure of turbulence in the boundary layer (Hashmonay, 1991)

Its importance there is no direct method available to determine the MH. The most common methods for determining the MH are the utilization of radio-sounding, remote-sounding systems, and parameterization methods.

The ceilometer, based on the lidar-technique is measures the aerosol concentration profile. Since in general aerosol concentrations are lower in the free atmosphere than in the mixing layer where most sources of aerosols are located, it can be expected that MH is associated with a strong gradient in the vertical back-scattering profile.

The mixing height can be determined by a wind profiler from the signal-to-noise ratio (SNR). The return signal is received primarily from the inhomogeneities of the radio refractive index. (Angevine, 1994)

The Vaisala single-lens ceilometer CT25K (Eresmaa, 2006) measures the optical backscatter intensity of the air at a wavelength of 905nm (near infrared). Its laser diodes are pulsed with a repetition rate of 5.57kHz. The lens has a focal length of 377mm and an effective diameter of 145mm. The laser beam full divergence and field-of-view divergence of the receiver are 1.4mrad each. Because of the monostatic optical system and the small divergence multiple scattering effects are negligible and the Mie scattering with scattering angles between 179.9° and 180.1° is dominant.

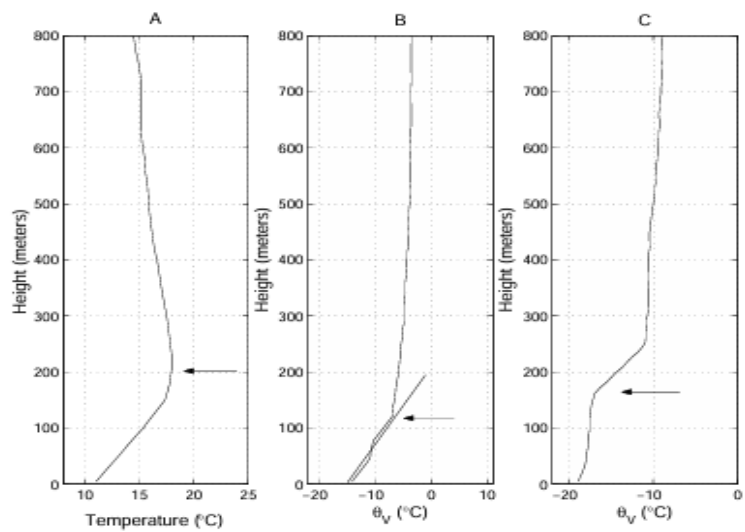

*Figure 21. Three ways for determining the reference mixing height from temperature profiles: (a) the height of the surface inversion (b) virtual potential temperature non-linearity and (c) strong winds–sharp virtual temperature increase above MH (Source: Vaisala Ceilometer Manual)*

In our study, we have used the data from the Vaisala CL31 ceilometer situated in PRL Ahmedabad (23.03571, 72.54378) that observes the backscatter using LIDAR technology. We have acquired the data from 10-05-24 to 14-05-24 encapsulating the occurrence of the sandstorm on 13$^{th}$ May. We observed the amount of backscattering in a period of 4 days and visualized the results using MATLAB. The cloud presence and boundary layer variation in the following days were also studied to give an idea about the presence of aerosols and clouds during the dust storm.

### 3.1.1.3    SABER:

SABER (Sounding of the Atmosphere using Broadband Emission Radiometry) is a 10-channel infrared radiometer that is one of four instruments on the NASA TIMED (Thermosphere-Ionosphere-Mesosphere Energetics and Dynamics) satellite mission to study the structure, energetics, chemistry, and dynamics of the Earth's mesosphere and lower thermosphere. Each filter of SABER has a particular filter over its detector to pass a particular spectral region to the sensor (Esplin, 2023).

The SABER experiment investigates the thermal structure and energy balance of the MLT. The energy budget primarily includes heating caused by solar radiation absorption and exothermic chemical reactions, which degrade solar energy to heat, and infrared radiative cooling. Mlynczak and Solomon (1993) provide a detailed analysis of solar energy deposition, non-cooling radiative losses, and heating through exothermic chemical reactions.

To properly understand the radiative energy budget of the MLT, it is necessary to observe kinetic temperature (T), ozone ($O_3$), water vapor ($H_2O$), carbon dioxide ($CO_2$), nitric oxide (NO), atomic oxygen (O), and atomic hydrogen (H) (Mlynczak, 1996, 1997). These measurements reveal the vertical contours of The SABER data that are being used to examine long-term changes and trends in temperature, water vapor, and carbon dioxide. A tacit, central assumption of these analyses is that the SABER instrument radiometric calibration does not change with time; that is, the instrument is stable. SABER stratospheric temperatures and those derived from Global Positioning System Radio Occultation measurements are compared to examine SABER's stability (M.G, et al., 2020).

The SABER instrument and preliminary calibration performance are described in Russell et al. (1999). The single most important decision made in the development of SABER was to commit to producing the most accurately calibrated instrument possible for the available resources. (Russell, 1999)

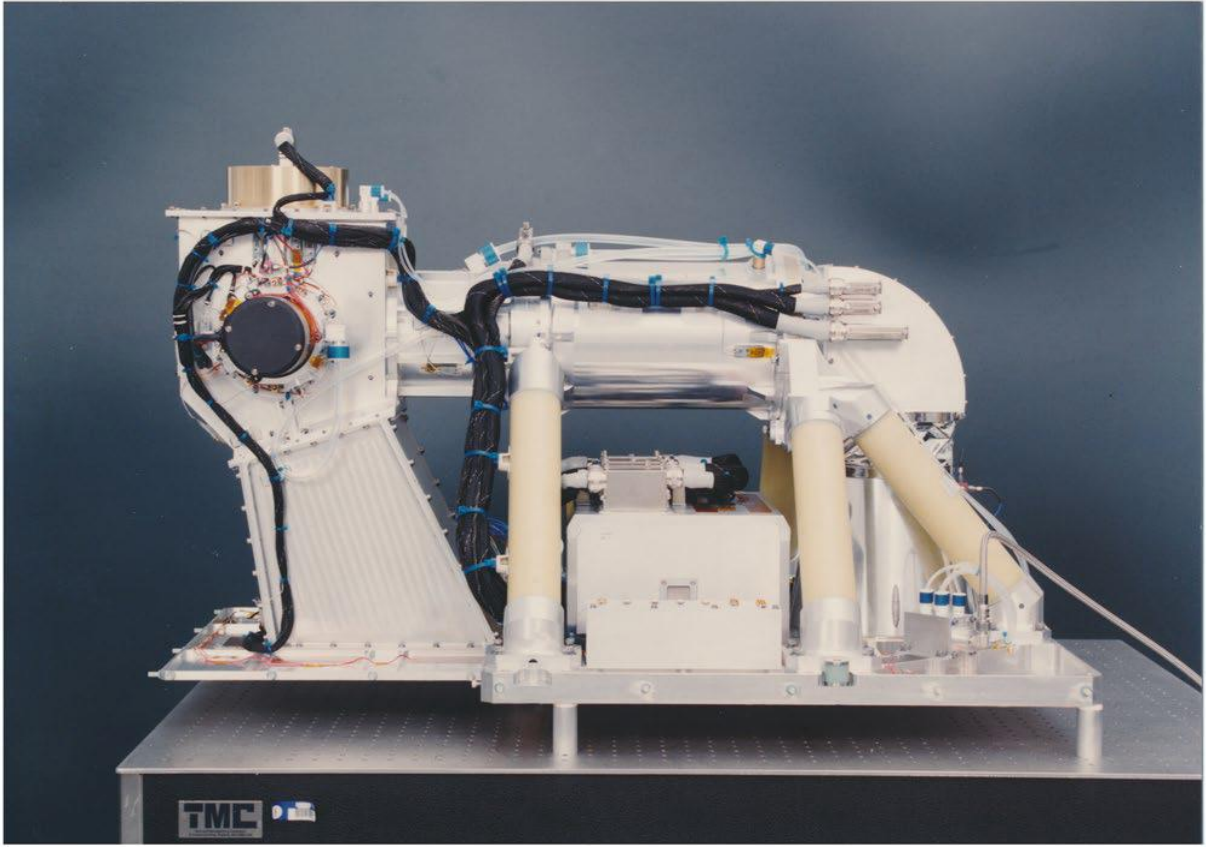

*Figure 22. SABER satellite before it was covered with multi-layer insulation (Esplin, 2023)*

The 10 channels of the satellite can detect radiation of a unique wavelength, collect the data, amplify it, and send it for analysis.

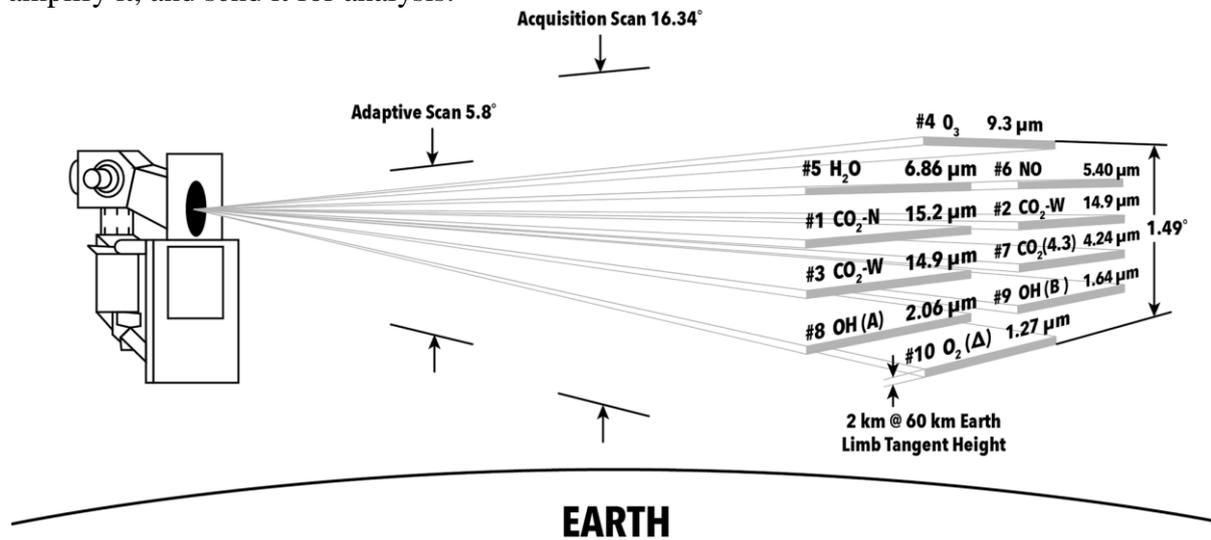

*Figure 23. Instantaneous Field of View (IFOVs) of the 10 SABER detectors on the atmosphere (Esplin, 2023).*

Recent studies have analysed trends in temperature (Garcia et al., 2019), carbon dioxide (Yue et al., 2015; Rezac et al.,2018), and water vapor (Yue et al., 2019) using SABER data. A tacit but central assumption of these analyses is that the SABER instrument has remained stable; that is, its absolute radiometric calibration is unchanged, over the life of the instrument; or, that any changes with time are significantly smaller than the observed trends.

In our study, we have used SABER data to find the O$_3$ mixing ratio in two different bands (1.27μm and 9.3μm), the H$_2$O mixing ratio, and the variation of temperature and plotted them with the variation of altitude in the y-axis. Data from 2 places having different geographical and atmospheric conditions are taken and compared to see the variation in the SABER data with changes in time and climate conditions.

### 3.1.1.4    ERA5:

ERA5 represents the fifth generation of atmospheric reanalysis developed by the European Centre for Medium-Range Weather Forecasts (ECMWF), covering the period from 1950 to the present. This reanalysis is produced in an operational mode, available to the public with a five-day delay, and is projected to continue for the next 5–10 years. ERA5 is part of a suite of climate data products curated by the European Union's Copernicus Climate Change Service (C3S), which is hosted at ECMWF (Thépaut et al., 2018).

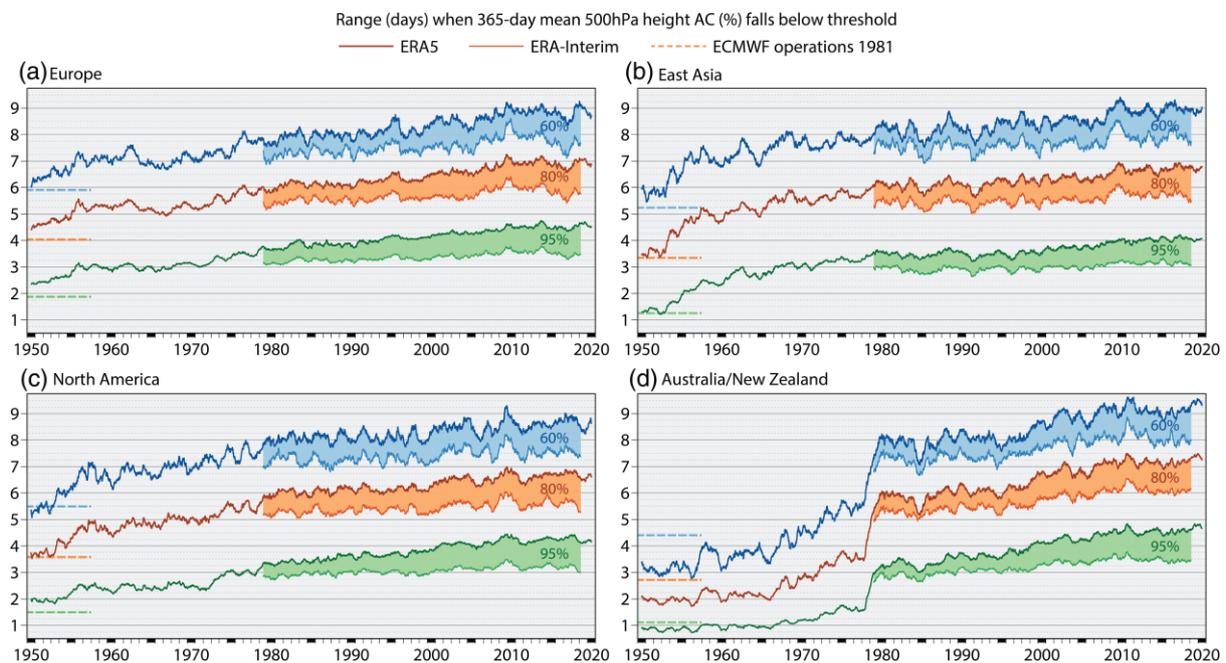

*Figure 24. Range (days) at which running 365-day mean anomaly correlations of 500 hPa height forecasts at 0000 and 1200 UTC from 1950 to 2020 reaches 95% (green), 80% (orange), and 60% (blue) for (a) Europe, (b) East Asia, (c) North America and (d) Australia/New Zealand. Also shown (dashed) is the average skill of ECMWF operational forecasts for 1981. Heavy lines denote ERA5; thin lines denote ERA-Interim. Shading denotes the difference between ERA5 and ERA-Interim during the period for which both are available (1979–2019) (Hersbach H, 2020)*

ERA5 delivers hourly estimates of the global atmosphere, land surface, and ocean waves at a high resolution (HRES) of 31 km, a notable improvement from its predecessor, ERA-Interim, which had a resolution of 78 km. Additionally, it provides uncertainty estimates through an underlying ten-member ensemble of data assimilations (the EDA product) at half the resolution with 3-hourly output. The main features of ERA5, along with comparisons to contemporary reanalysis for the period 1979–2019 (referred to as ERA579→), are discussed in Hersbach et al. (2020). This article details the characteristics of the recently completed segment of ERA5, spanning January 1950 to December 1978 (ERA5→79), supplementing and updating Hersbach et al. (2020) by contextualizing trends and diagnostics for the earlier period within the complete ERA5 record.

Reanalysis is crucial for a broad range of applications, from inter-governmental assessments of global climate change (Stocker, 2013) to specific and unique use cases necessitating precise local weather representations. Consequently, reanalysis strives to provide homogeneity and accuracy in representing global and regional climate variables over multi-decadal timescales, as well as accuracy in depicting synoptic-scale events at sub-daily temporal resolutions. For large-scale applications, key aspects of reanalysis include accurately representing the evolution of the mean state of the atmosphere (including thermodynamic, dynamical, and radiative variables) and extremes.

ERA5 data is freely accessible through the Copernicus Climate Data Store (CDS). Users can download data via the CDS web interface or use APIs for automated data retrieval, making it convenient for researchers and practitioners to obtain the needed information (Copernicus Climate Change Service, 2023). The extensive range of variables and high-resolution data make ERA5 suitable for various applications, including climate research, weather forecasting, hydrology, renewable energy, environmental monitoring, etc.

Despite its advancements, ERA5 is not without limitations. Certain regions, especially over the oceans and polar areas, suffer from sparse observational data due to the resolution of the model, leading to potential biases and uncertainties in the reanalysis (Simmons, et al., 2020).

### 3.1.1.5 MODIS:

MODIS stands for Moderate Resolution Imaging Spectroradiometer. It was launched by NASA on December 18, 1999, equipped on the TERRA platform.

The Moderate Resolution Imaging Spectroradiometer (MODIS) collects data in 36 spectral channels (0.4μm to 14.4μm spectral width) with global coverage every 1 to 2 days. MODIS data's very broad spectral range allows it to be used in a variety of research, including vegetative health, changes in land cover and land use, oceans and ocean biology, sea surface temperature, and cloud analysis. It is also widely used for monitoring fires, natural dangers, and oil spills. MODIS data products are available in real-time or near real-time, which is a significant feature. Direct broadcast stations worldwide download raw MODIS data in real-time directly from the satellite, while NASA's Land, Atmosphere Near Real-time Capability for EOS (LANCE) supplies numerous MODIS products within three hours of the satellite.

NASA has several datasets from a variety of sensors linked to air quality that are freely and openly available via Earth-data Search and the agency's websites. The Moderate Resolution Imaging Spectroradiometer (MODIS) instrument on NASA's Aqua and Terra satellites is a valuable data source for air quality studies because it measures aerosol and other data outputs. Terra also has the Multi-Angle Imaging Spectroradiometer (MISR), which measures aerosol particles.

We have used MODIS-Terra and Aqua to collect data on various parameters around Ahmedabad and Mount Ruang.

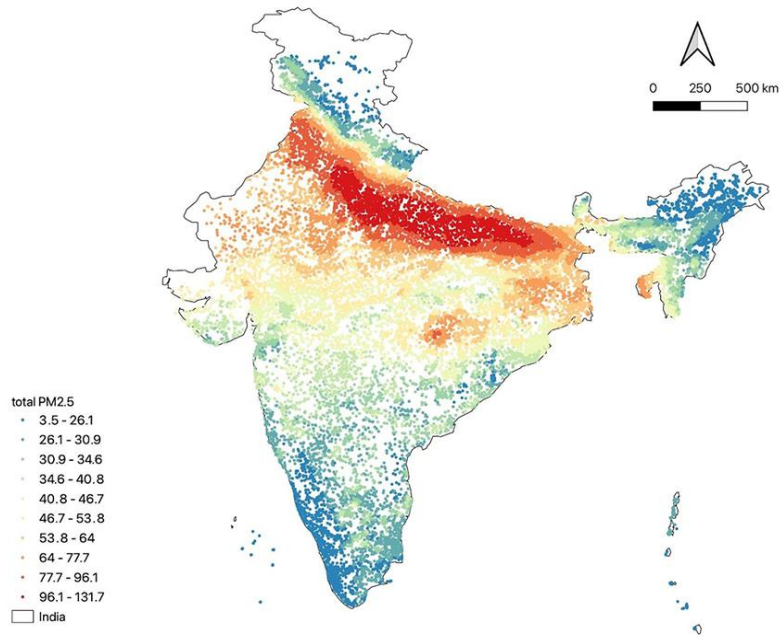

*Figure 25. This image shows average PM2.5 levels across India from 2010 to 2015. Blue and green colours indicate lower levels of particulates in the air; yellow and red colours indicate areas with high levels of pollutants, such as along the foothills of the Himalayas. Credit: Priyanka D'Souza (2023).*

### 3.1.1.6 MERRA-2:

The Modern-Era Retrospective Analysis for Research and Applications, Version 2 (MERRA-2) provides re-analysis data starting from 1980. It was introduced to replace the original MERRA dataset because of the advances made in the assimilation system. It also uses NASA's ozone profile observations that began in late 2004. Additional advances in both the GEOS model and the GSI assimilation system are included in MERRA-2. Spatial resolution (0.5° x 0.625°) remains about the same (about 50 km in the latitudinal direction) as in MERRA. The data is available at the official website [MERRA-2 (nasa.gov)](MERRA-2).

Retrospective analysis (reanalysis) data products are based on the assimilation of a vast number of in situ and remote sensing observations into an atmospheric general circulation model (AGCM) and provide global, sub-daily estimates of atmospheric and land surface conditions across several decades (Reichle, et al., 2017). The MERRA-2 is capable of significantly more assimilation, with almost 5 million observation assimilations in a 6-hour cycle, compared to 1.5 million for MERRA. MERRA-2 uses an automated bias correction scheme for the assimilation of most satellite radiance observations. Bias estimates for individual sensor channels are represented by a small number of predictors, which can depend on the atmospheric state, the radiative transfer model, and the sensor characteristics (Gelaro, et al., 2017).

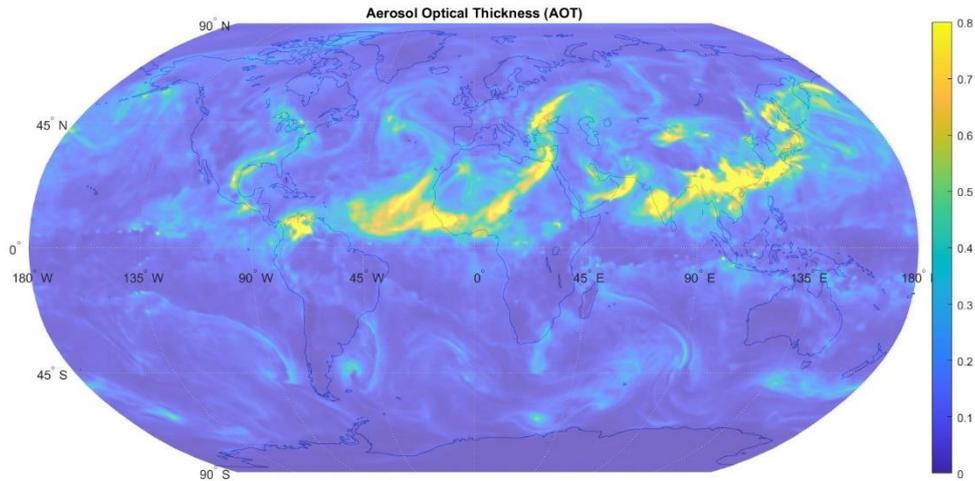

*Figure 26. This image shows the global AOT variation in April 2024. MERRA-2 reanalysis data is used to create the distribution. The yellow areas show a high concentration of aerosol particles.*

In this study, we have used the MERRA-2 data to observe aerosol optical thickness (AOT) of various places including Ahmedabad, Mount Ruang, and the global AOT distribution of April 2024. The data is publicly available on the website for download and use.

### 3.1.1.7 Giovanni:

Giovanni is a NASA Goddard Earth Science Data and Information Services Center (GES DISC) Distributed Active Archive Center (DAAC) web application that provides a simple, intuitive way to visualize, analyze, and access Earth science remote sensing data, particularly from satellites, without having to download the data. Giovanni webpage is free to use and interactive in plotting data from various satellites without having to download the datasets. We have used this website to get a general view of the aerosol distribution over India from the date 12-05-24 to 18-05-24 and visualize the aerosol optical depth (AOD).

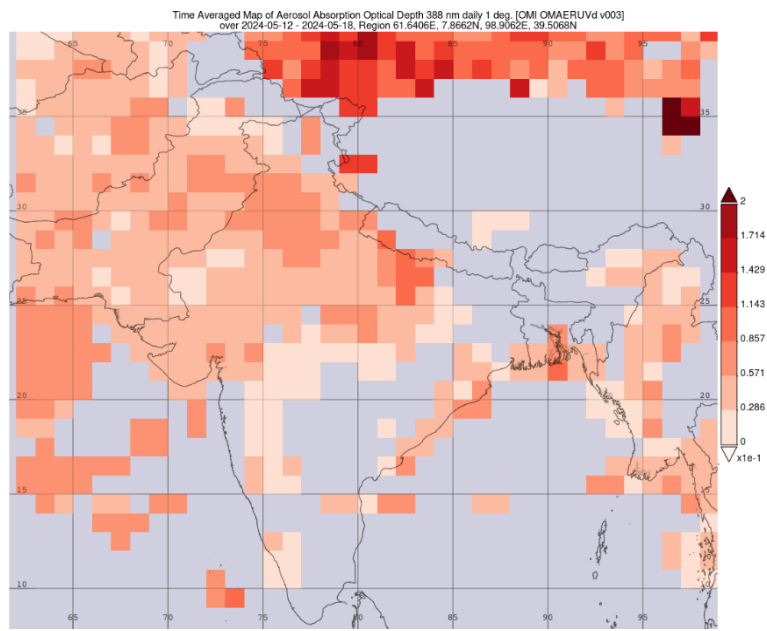

*Figure 27. AOD distribution over India in May 2024 (Giovanni, NASA Earth data)*

Giovanni provides a user-friendly interface for exploring datasets related to the atmosphere, oceans, and land, facilitating the study of climate change, weather patterns, and environmental phenomena. Researchers and educators utilize Giovanni for its capability to generate time-series plots, maps, and other graphical representations of data, making complex scientific information more accessible and comprehensible.

### 3.1.1.8   INSAT-3DR:

MOSDAC (Meteorological and Oceanographic Satellite Data Archival Centre) it's an important repository managed by the Indian Space Research Organisation (ISRO). MOSDAC archives data from satellites such as INSAT (Indian National Satellite System) and Megha-Tropiques, among others. This data includes meteorological observation, oceanographic data, and information on atmospheric conditions.

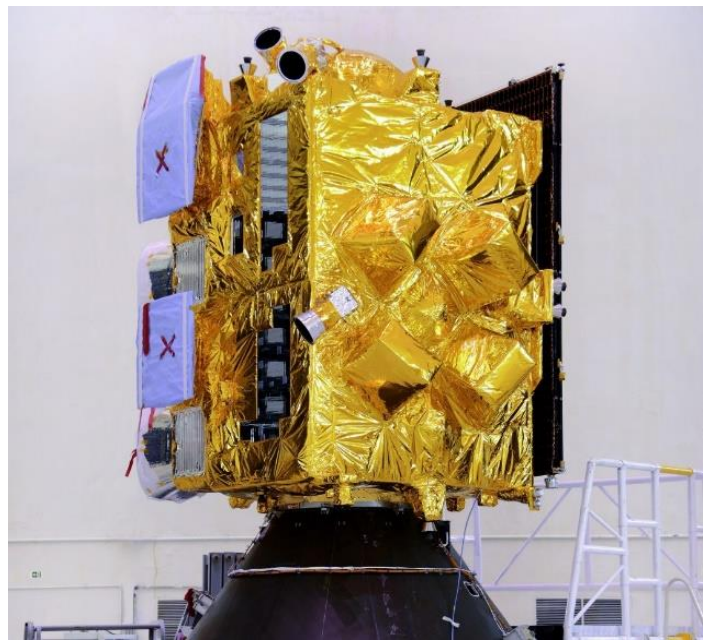

*Figure 28. INSAT-3DR. An advanced meteorological satellite. Launched on September 8, 2016. It is a part of the INSAT series and enhances the capabilities of its predecessor. INSAT-3D and INSAT-3DR satellite data are used in forecasting/monitoring weather. (Credits: ISRO)*

INSAT -3DR is a multi-spectral imaging system capable of capturing images in six different spectral bands, including visible, shortwave infrared, middle infrared, water vapor, and two thermal infrared bands. The orbital positioned in a geostationary orbit at 74 degrees East longitude, providing continuous coverage of the Indian subcontinent and surrounding regions. Imaging in middle infrared band to provide night time picture of low cloud and fog and two thermal infrared bands for estimation of sea surface temperature (SST) for better accuracy. High spatial Resolution in the visible and thermal infrared bands.

INSAT-3DR uses an omnidirectional antenna for the downlink payload data. The data relay transponder accepts UHF signal (450 MHz) from various data collection platforms on earth and downlinks the collected data in C-Band at 4500 to 4510MHz. Its dimensions of 2.4m × 1.6m × 1.5m, and it has a liftoff mass of 2211kg, which includes about 1255kg of propellant (dry mass of 956 kg). The satellite has 3 payloads Meteorological (MET) – IMAGER and SOUNDER.

In this study, we have used MOSDAC to gather high-level wind vector data for the study locations Ahmedabad and Mt. Ruang on the dates of concern. The wind vector figures are taken from the free-to-access website.

### 3.1.1.9 Automatic Weather Station (AWS)

An automatic weather station (AWS) is an automated version of the traditional weather station, either to save human labour or to enable measurement from remote areas.

In our study, we are using the WXT536 model which measures pressure, temperature, humidity, rain, wind speed, and wind direction that uses ultrasound to determined horizontal wind speed and direction. The ultrasonic pulses are sent between the transducer and the time it takes for the pulses to travel between them is measured.

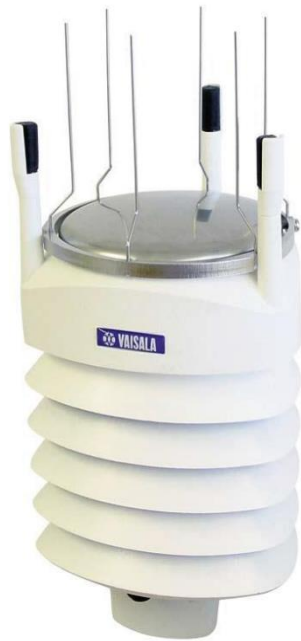

*( Fig.28.b. Vaisala WXT536 AWS (Source: https://up1.goepe.com/)*

Its small size makes it ideal for quick short-term deployments. The array of three equally spaced transducers on a horizontal plate is a Vaisala specific design.

The WXT536 uses a capacitive sensor to detect raindrops. The sensor measured the impact and size of raindrops to calculate the rainfall rate. The device can differentiate between liquid and solid precipitation to some extent ensuring more accurate measurement.

A capacitive barometric pressure sensor measured atmospheric pressure. The sensor consists of a diaphragm and a reference vacuum. Changes in atmospheric pressure cause the diaphragm to deflect changing the capacitance which can convert into a pressure reading.

The temperature is measured using a precision thermistor. The thermistor changes resistance with temperature and it is measured and converted to a temperature. The Temperature range is -52°C to 60°C with a measurement accuracy of ±0.3°C. The measurement range for humidity is 0-100% with an accuracy of ±3%.

We are using the AWS to collect data of temperature variation with time during the dust storm.

# 4 CHAPTER 4

## 4.1 OBSERVATIONS AND RESULTS:

### 4.1.1 Observing the Atmosphere of Ahmedabad Using Sunphotometer:

The sun photometer was used in this project to take observations of ozone, water vapor, and AOT (aerosol optical depth) over the region of Ahmedabad (23.03571, 72.54378). Readings were taken from the sunphotometer from 15/05/24 to 19/06/24. In four readings each day, spanning two hours i.e. 10:00 am, 12:00 pm, 2:00 pm, and 4:00 pm. We obtained a similar pattern in almost all graphs and no significant variation was observed except on 15th May when there were some observed clouds. Graphs showing the changes in Ozone, water column depth, and AOT (aerosol optical depth) were plotted in a time-series format.

The data collected from the MICROTOPS II sunphotometer were carefully taken to minimize the errors, as reported previously in the studies the bias is relatively small.

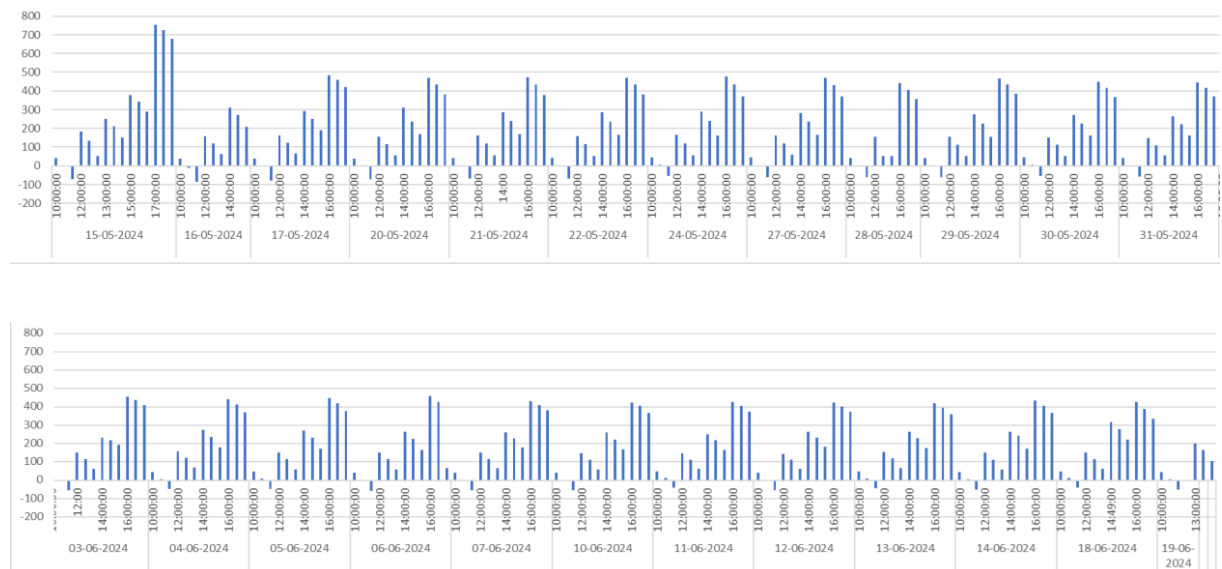

*Figure 29. This figure shows the variation of Ozone (in DU) during daytime observed by MICROTOPS II sunphotometer over a period of a month. On x-axis we have the datetime and on y-axis we have the ozone concentration in Dobsons Unit (DU). It shows a pattern where the concentration increases gradually as the day progresses.*

Ozone levels can increase as the sun undergoes cycles, becoming more active and emitting higher amounts of UV rays. Ozone is affected by the changes in sunlight and temperature that occur during different seasons. The concentration of ozone was greater in the graph due to cloudy weather on 15 May 2024. while other days showed little variation. In practice, MICROTOPS II ozone observations are more consistent and agree better with Dobson and Brewer's ozone measurements when the sky is free of haze. (Morys, 2001)

The aerosol scattering when ozone is computed from two pairs of wavelengths. For example, moderate haze does not generally affect ozone measurements by more than a few percent and much less under usual conditions (Basher, 1997) Solar UV passing through the atmosphere can be modulated as much by haze in the troposphere as by ozone in the stratosphere.

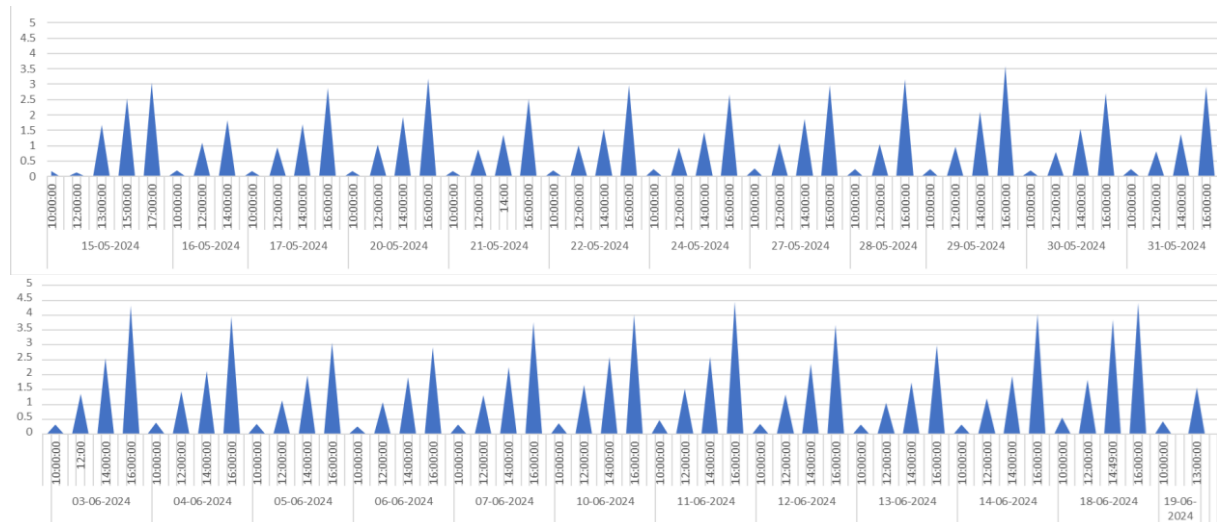

*Figure 30. The figure shows the gradual change in water column depth (in cm) in y-axis with the date time in x-axis. The pattern shows increase in the column depth and peaks around 4pm. It is a stable pattern noticed throughout the month of observation (May-June).*

From figure.30 we can see the water column depth grows deeper over time, mainly due to rising temperatures that boost evaporation and the atmosphere's ability to hold more water vapor. The water column depth in the sun photometer deepens over time, primarily because the temperature is on the rise, which in turn stimulates evaporation and enhances the atmosphere's capacity to hold more water vapor. The ratio of the 940 nm and 1020 nm channels is the measurement of total column water vapor. Cumulus clouds were present, the column water vapor was moderately high. (Morys, 2001)

The water vapor measurement is based on a pair of radiometric measurements in the near-infrared. The 940 nm filter is located in a strong water vapor absorption band, while the 1020 nm filter is affected only by aerosol scattering. The optical thickness of haze is often related to the column amount of water vapor, two of the five channels in Microtops and MICROTOPS II have been traditionally devoted to the measurement of column water vapor. This measurement also provides the optical thickness at 1020 nm, that is at an atmospheric window free of gaseous absorption. (Morys, 2001)

The attenuation of sunlight caused by water vapor has been studied for almost a century. The calibration technique used for Microtops was developed by Reagan et al. (1987) and further tested by Michalsky et al. (1995). The water vapor measurement is based on a pair of radiometric measurements in the near-infrared. The 940 nm filter is located in a strong water vapor absorption band, while the 1020 nm filter is affected only by aerosol scattering

Column water vapor is measured each time MICROTOPS II measures column ozone. As the emphasis with MICROTOPS II and its predecessors has been the measurement of total ozone, the column water vapor measurements by MICROTOPS II have yet to be fully analyzed. (Morys, 2001)

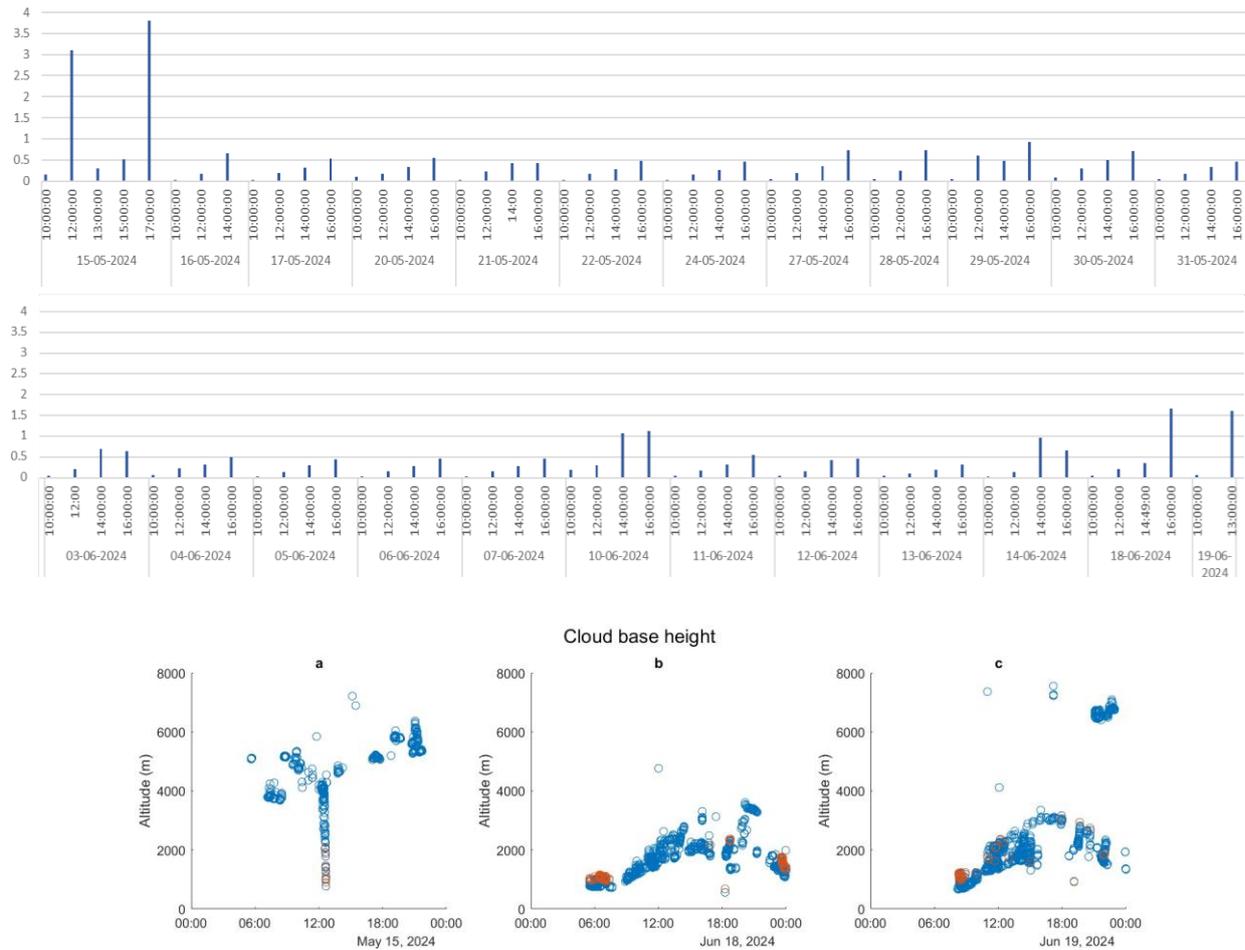

*Figure 31. The figure shows the plot between Aerosol Optical Thickness (AOT) on the y-axis with time on the x-axis in May-June. The AOT shows a standard pattern with occasional peaks at times when the concentration of aerosols is high. On the dates with high AOT presence of the clouds can be seen from ceilometer plot below.*

The Fig. 31 shows a gradual increase in AOT as the day progresses and decrease in the afternoon. During rush hours in urban areas and dust storms, AOT values tend to be higher. In the summer AOT values are also increased due to biomass burning, dust, and the presence of clouds. The peaks are seen when there is the presence of clouds or dust carried by wind.

Readings vary even when there is no proper sun radiation. The maximum AOT value decreased with the increase in the wavelength for all wavelengths, similar to the mean of the wavelength this result is in agreement with the Angstrom law (the smaller the particles, the larger the exponent).

(Gong, Evalution of maritime aerosol optical depth and precipitable water vapor content from the Microtop II Sun photometer, 2018)

### 4.1.2 Study of the Dust Storm Over Ahmedabad on 13th May 2024:

A dust storm engulfed the city of Ahmedabad on 13th May 2024 drastically reducing the visibility and affecting the atmosphere. In this report, we discuss the impact of the dust storm on the atmosphere, the boundary layer, and local weather.

Ahmedabad (23.02°N, 72.57°E) is a sandy and dry area located at 53 meters above sea level. Dust storms are generally witnessed in the pre-monsoon season in March-May (Chakravarty, 2021) which can be seen in the Fig. 32.

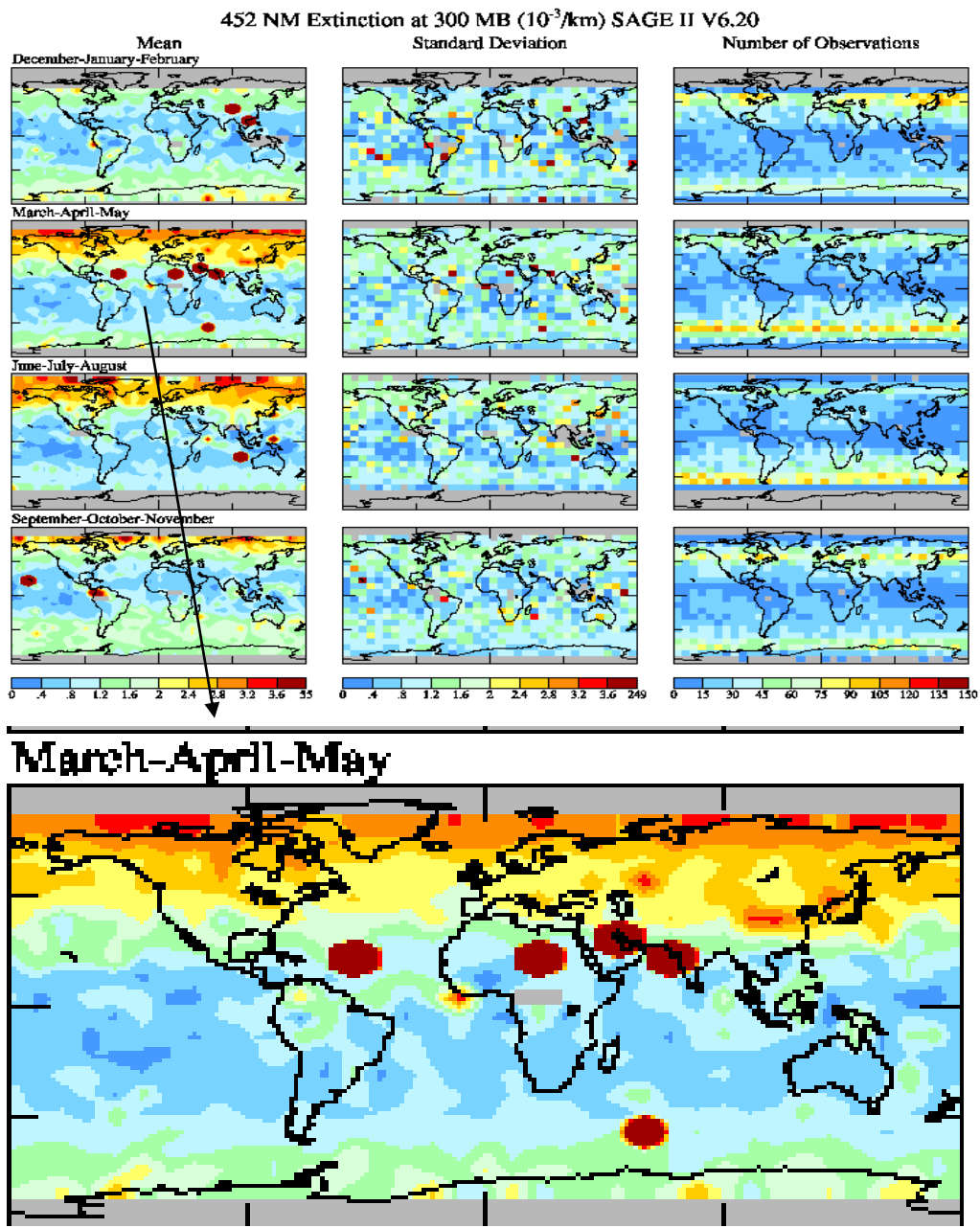

*Figure 32. Seasonal AOT variation using GISS (SAGE-II) data from 1985-2003. Here we can see a high concentration of AOT near the Gujrat region during the month March-April-May.*

The high concentrations of aerosol are due to the presence of dust and other CNN (Cloud Condensation Nuclei) in the atmosphere. These aerosols absorb the long-wave radiation reducing the surface temperature and increasing the humidity (Saha, et al., 2022 & Satheesh et al., 2005). This was confirmed by observing the longwave flux from the top of the atmosphere using the ERA-5 reanalysis data.

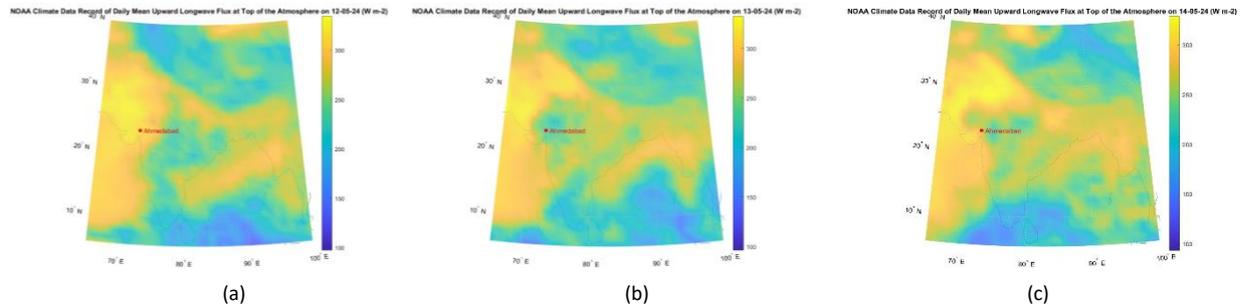

*Figure 33. Daily mean upward longwave flux on the dates 12, 13, and 14 May using ERA-5 data. (a) There is not much obstruction to the upward flux. (b) Around 1.5 times decrease in the flux on the day of the dust storm due to the high level of aerosols. (c) The aerosol is distributed and absorbs the longwave radiation wherever present in the atmosphere.*

There is also a decrease in the temperature which is visible in Fig.34 below. The data from AWS located in PRL, Ahmedabad is used to show a decrease in the temperature during the event. A significant increase in the relative humidity due to an increase in the CNN can also be seen from the plot as observed previously by Saha et al. (2022).

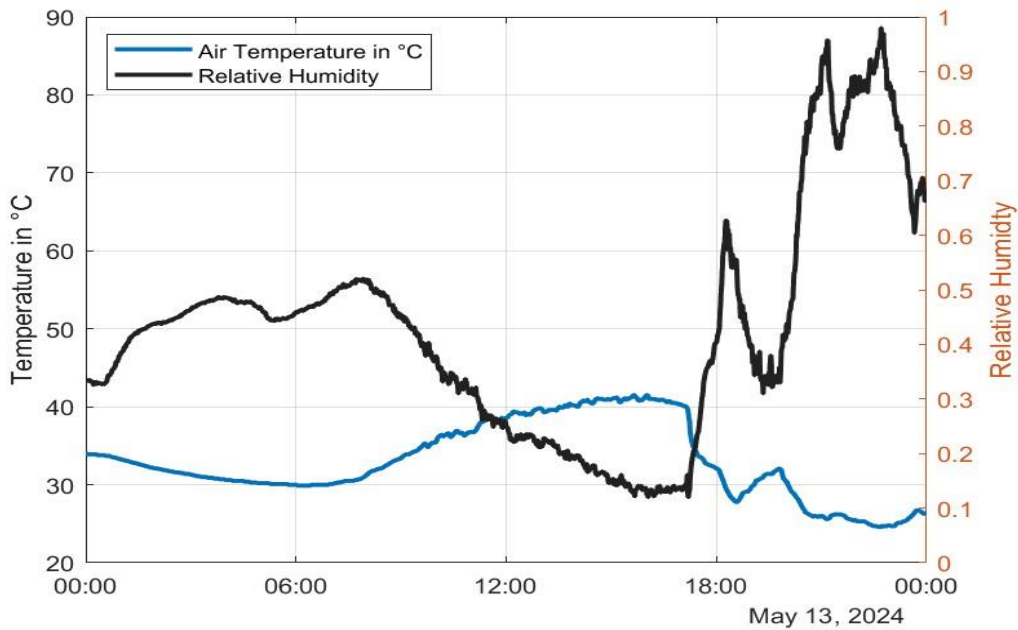

*Figure 34. Variation of temperature and relative humidity on 13th May with respect to time using the AWS in PRL. This shows a time series plot of the variables where sudden changes are seen near 6pm. The temperature decreases rapidly and relative humidity shows a peak at that time.*

In Fig. 34 the temperature can be seen to decrease from above 40 ºC at 4 pm to 28 ºC around 6 pm, the time when the dust storm peaked. The sudden decrease in the temperature was due to the aerosol cooling by the dust storm as well as the rain accompanying it. The relative humidity drastically increased after 5 pm to reach a peak at 65% at 6 pm and can be seen fluctuating in a short period of time.

Due to this variation in temperature, increasing AOT, strong wind, etc. the atmosphere is unstable during the dust storm. The boundary layer does not follow a diurnal cycle and breaks down. The destruction of the boundary layer due to rapid mixing and influx of aerosols into the boundary layer creates a dynamic atmosphere (Singh et al., 2022). The study of backscatter data from the ceilometer present at the PRL, Ahmedabad shows a similar finding.(Fig. 35)

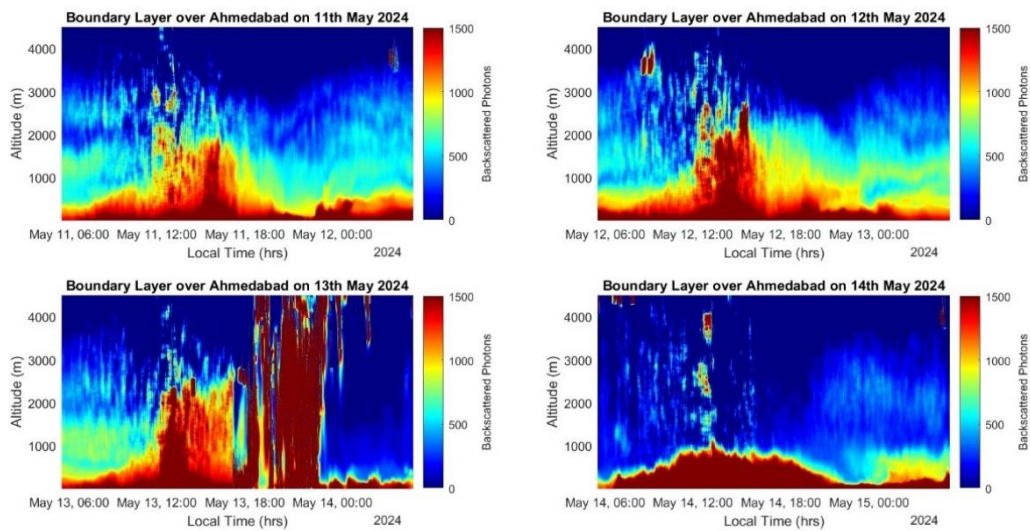

*Figure 35. Backscattering data of aerosols observed by the VAISALA Ceilometer in PRL. The Boundary Layer of four days is shown with dates on the above.*

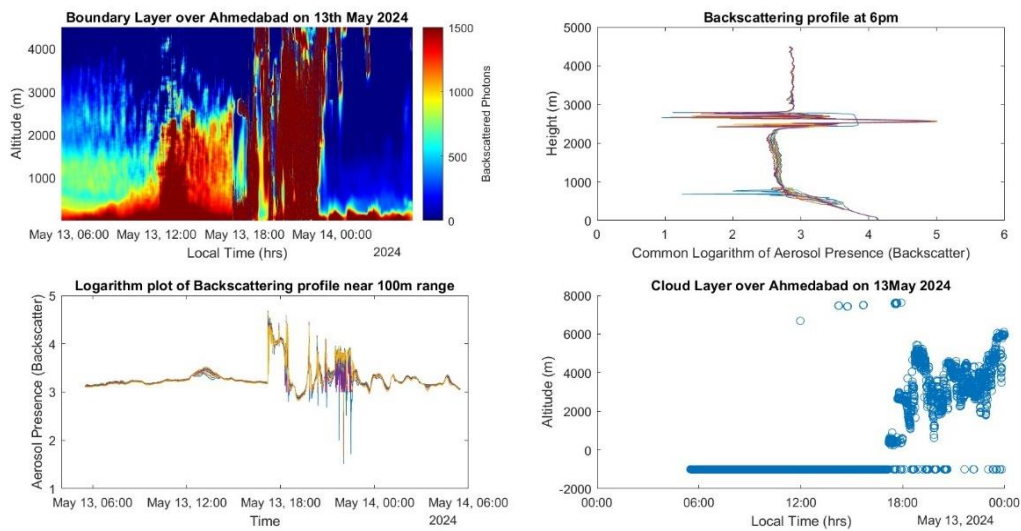

*Figure 36. Presence of clouds and aerosols in the atmosphere on 13$^{th}$ May. Top left shows the boundary layer variation. Top right shows the logarithm of the backscattering coefficient with height around 6pm. Bottom left is the BS profile with time near the surface up to a height of 100m. Bottom right shows the presence of clouds.*

In Fig.35 the BL (boundary layer) follows a standard diurnal variation on 11th and 12th May. But on 13th May, the day of the dust storm event, we can see the increase in backscattering just before the dust storm showing rapid mixing and entrainment. Around 6 pm the BL was completely destroyed due to the high concentration of aerosol and the presence of cloud. This causes the backscattering to drastically increase as seen in the figure. The dynamics of the atmosphere during the dust storm can be visualized from the Fig. 36 where backscatter and cloud data are shown. There was high backscattering near the altitude of 2500m due to the presence of clouds but backscattering was observed near the 100m range from the surface which is due to the presence of dust.

The dust in the lower region could have acted as CNN which led to the increased formation of clouds during the event (Wallace & Hobbs, 2006).

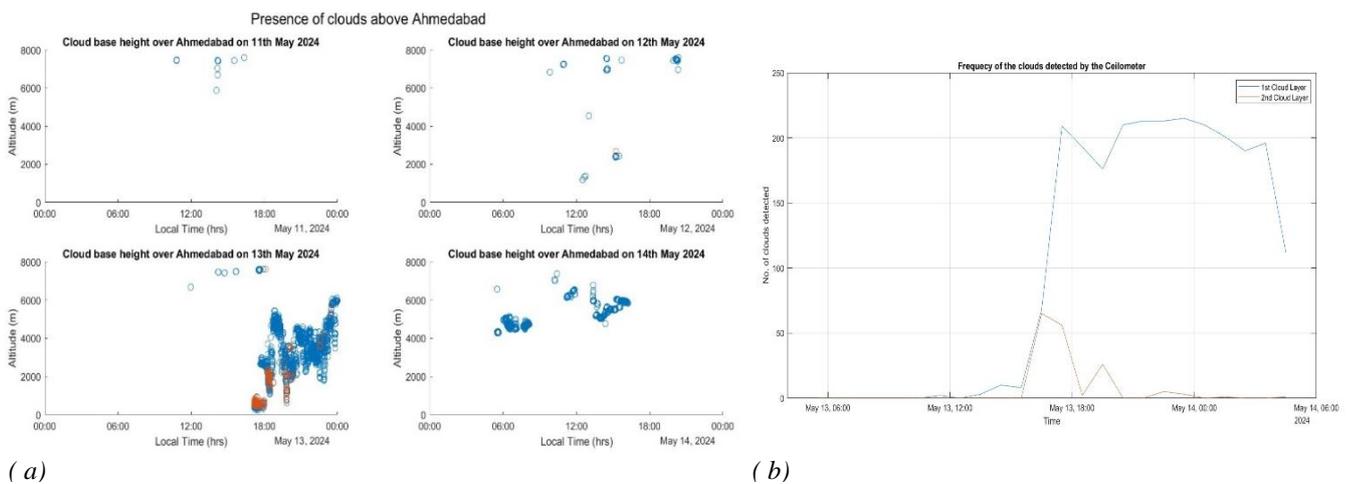

*( a )*                                                                                                      *( b )*

*Figure 37. (a) Cloud data of four days using ceilometer, blue dot represent clouds of 1st layer and red dots are clouds of 2nd layer. (b) Cloud frequency of both layers on 13th May with time on x-axis.*

Fig. 37 shows an increase in the amount of clouds after the event. Clouds of two layers can be seen, one layer with a low altitude of 1000m and another layer at an altitude of 6000m. Fig. 37(b) shows the frequency of clouds detected where the peak reaches 215. The rain followed the dust storm. The massive amount of clouds and rain caused a reduction in surface temperature the next day.

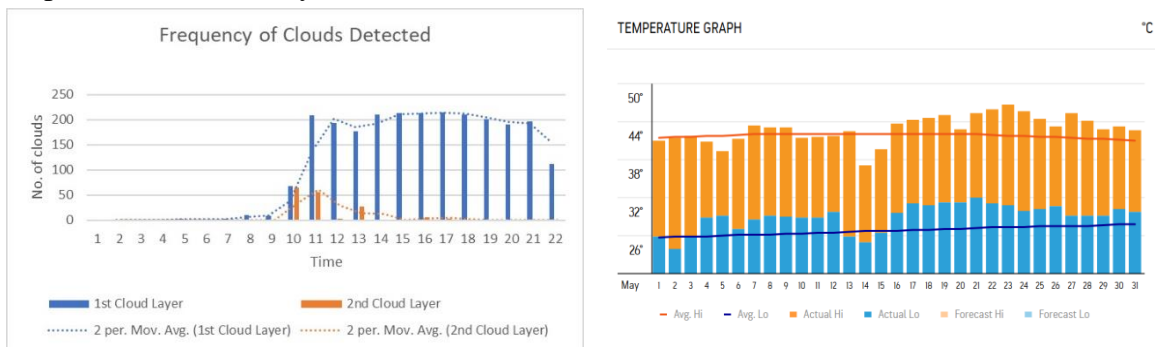

*Figure 38. Left shows histogram of the frequency of clouds present in the atmosphere on 13th may. Right shows a decrease in diurnal temperature decreasing the next day of the event.*

### 4.1.3 Study of the Volcanic Eruption in Mt. Ruang, Indonesia on 17th April 2024:

Ruang is a small volcanic island in the Sangihe Islands arc in North Sulawesi, Indonesia (2.305, 125.365). Late on April 16, 2024, the mountain unleashed a series of explosive eruptions that, at times, sent plumes of ash and gas billowing high into the stratosphere. In this report, we have observed the effect of the volcanic event on the atmosphere and its constituents, temperature, longwave flux, and general demography of the surroundings.

We have used Landsat-9-Operational Land Imager-II (OLI-II) to gather information about the demographics and clouds over Mt. Ruang before and after the event.

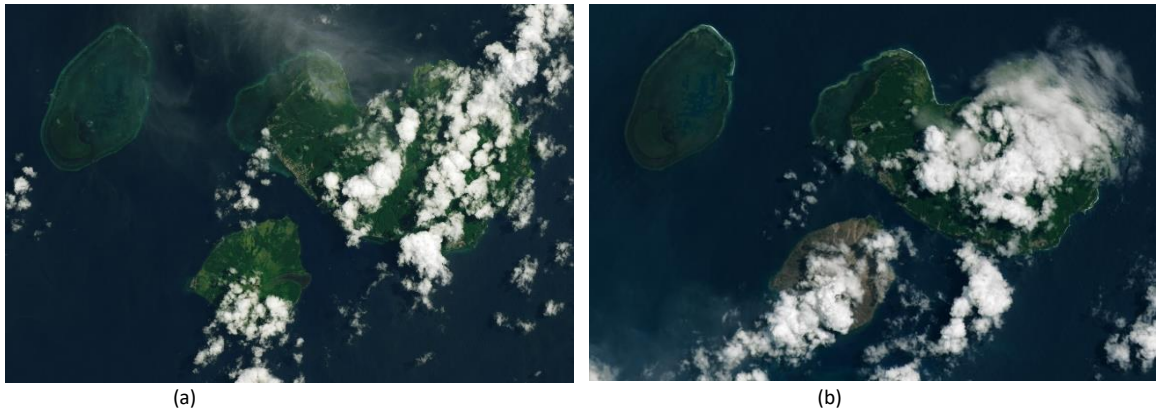

*Figure 39. OLI-II Satellite image of Mount Ruang in the bottom. (a) Image of Mt. Ruang before the volcanic eruption, on the date 12-04-2024. (b) Image taken on 20-04-2024, after the volcanic eruption, comparatively high amount of tephra and clouds can be seen.*

These explosive volcanoes release tephra to the troposphere increasing the concentration of aerosols which in turn increases the formation of clouds as seen in Fig. 39. Before the explosion a smaller number of clouds were observed but after it the vegetation died and a blanket of ash and clouds can be seen on the right. The formation of clouds due to aerosols acting as CNN causes a decrease in the temperature (Kelly & Sear, 1984). The presence of aerosol can further be seen quantitatively by long wave flux absorption.

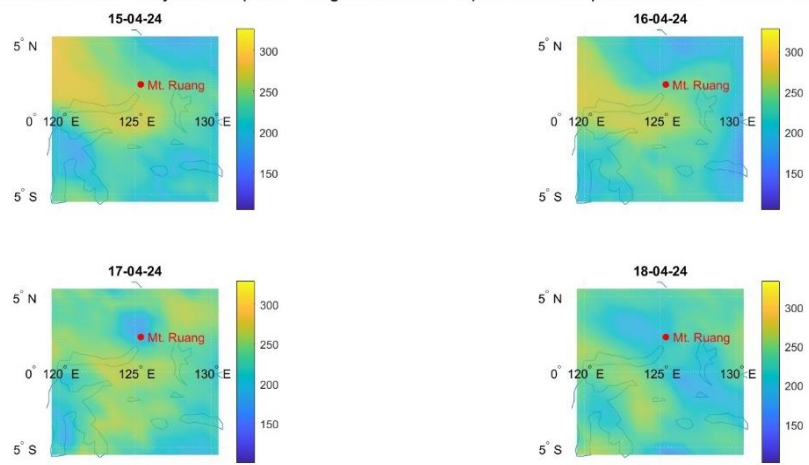

*Figure 40. Upward Longwave flux of four days including the event using the ERA-5 reanalysis data. On 17 April a decrease in the flux can be seen.*

Studying volcanic eruptions in the past such as Agung (1963), El Chichón (1982), and Mt. Pinatubo (1991), etc. there is temperature anomaly due to the events (Kelly et al., 1984; Mass & Portman, 1989; Robock, 2000). A decrease in temperature under the influence of clouds has been observed.

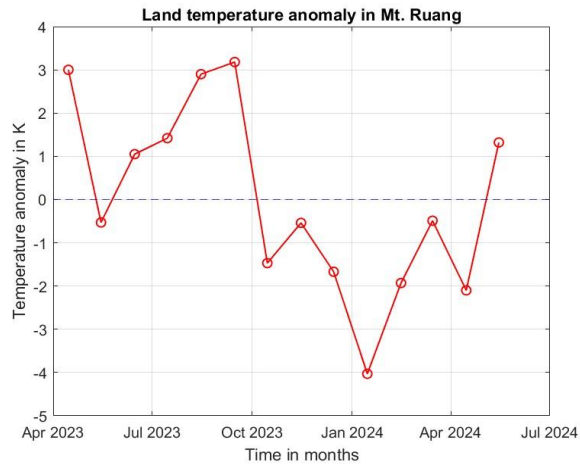

*Figure 41. Temperature anomaly near Mt. Ruang from Apr 2023 to Jul 2024 showing variation from the reference value. On April 2024 we can see a reduction of around 2 K from the standard temperature.*

Along with aerosol particles trace gases are also injected into the atmosphere through eruptions. Gases like $SO_2$, NO, $NO_2$, $HNO_3$, and HCl are released from it. Satellite observations showed the presence of HCl gas caused a reduction of the ozone concentration by up to 10% following the El Chichón and Mount Pinatubo eruptions (Coffey, 1996). Similar observations were carried out using the SABER satellite in this study showing a reduction of ozone concentration near stratosphere.

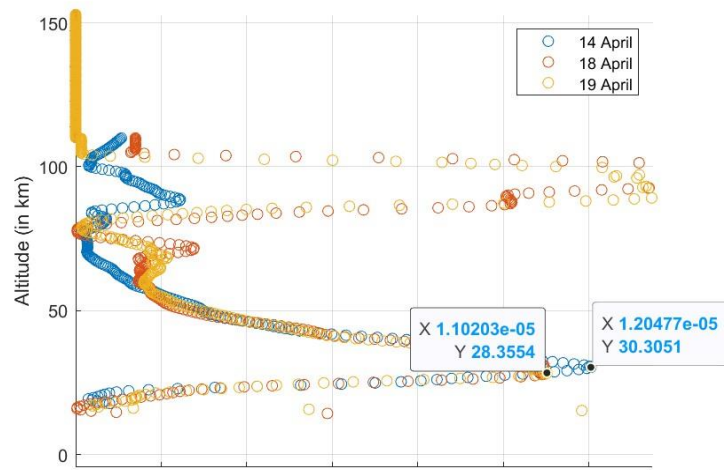

*Figure 42. Ozone mixing ratio using SABER data for three different days. A decrease in the mixing ratio is observed on 18 April after the eruption.*

Fig. 42 shows the Ozone mixing ratio profile with altitude before and after the eruption. Comparing the mixing ratio of 14 April with 18 April shows a decrease of the mixing ratio of 8.5% after the eruption. This can be because of the Chlorine present in the atmosphere due to volcanic eruption.

Another major constituent of volcanic gases is SO$_2$ (Sulphur Dioxide) which can be readily converted into sulphate aerosols. These cause acid rain when mixed with the atmosphere (von Glasow, 2009). In our study, a high concentration of SO$_2$ has been observed from the Ozone Monitoring Instrument (OMI-Aura). The SO$_2$ injected by the volcano was carried away by the west wind to further distances.

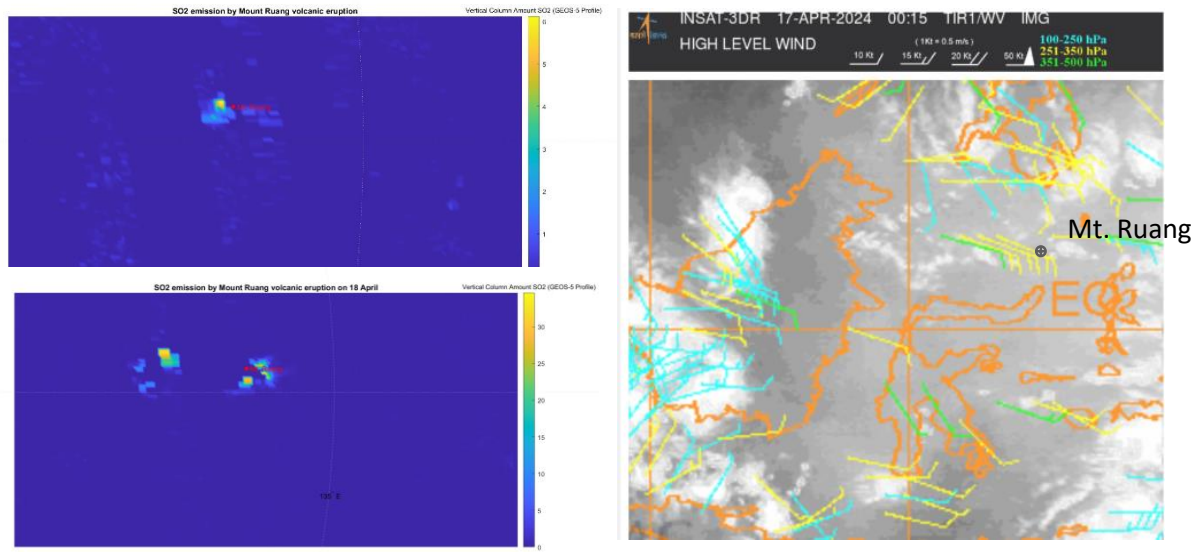

*Figure 43. Left: Vertical Column of SO$_2$ from OMI-Aura on 17 and 18 April showing the presence of SO$_2$ from the eruption. Right: Wind direction from INSAT 3DR showing west wind with high velocity on 17 April.*

Vertical column of SO$_2$ can be seen on the left where a high concentration of SO$_2$ can be seen near Mt. Ruang. The wind flow to west near the eruption place as seen on the right of Fig. 43 may have carried the atmospheric SO$_2$ towards west as can be seen on 18$^{th}$ April plot. Sulphate aerosols can be formed from the gas by chemical reaction. However, it takes around 2 weeks for SO$_2$ to convert into Sulfuric acid (von Glasow, 2009).

## 4.2 CONCLUSION:

In this study we explored the atmospheric changes during natural disasters, focusing on a dust storm in Ahmedabad and a volcanic eruption in Mount Ruang, which demonstrates the significant impact these events have on atmospheric composition, weather patterns, and overall environmental conditions.

### 4.2.1 Local Atmosphere Study:

We studied the local atmosphere using MICROTOPS-II sunphotometer from 15 May to 19 June for the parameters Ozone column, water column height, and AOT. Ozone levels show a similar pattern every day, having low concentration in the morning, reaching their peak in mid to late afternoon after exhaust fumes from morning rush hour have had time to react in sunlight. The higher ozone level is also due to increased solar radiation and higher temperatures the sun undergoes cycles, becoming more active and emitting higher amounts of UV rays. Ozone is affected by the changes in sunlight and temperature but it's not affected by the presence of clouds in the atmosphere.

The water column depth in the sun photometer deepens over time, primarily because temperature is on the rise, which in turn stimulates evaporation and enhances the atmosphere's capacity to hold more water vapor. A higher value of water column depth is observed due to increased humidity and precipitation, whereas lower values were recorded reflecting the drier atmospheric condition prevalent in the region.

The AOT values show a gradual increase due to the dynamics of the boundary layer by heating of the sun, which causes mixing of aerosols with higher atmosphere causing a higher vertical column. Then shows a decrease afternoon as the boundary layer begins to settle down. During rush hours in urban areas exhaust from vehicles may cause AOT values to be higher. The presence of clouds also causes a peak in the values of AOT. We have used 1020nm wavelength to calculate the AOT but the maximum AOD value decreases with the increase in the wavelength for all wavelengths due to different absorption profiles.

### 4.2.2 Dust Storm in Ahmedabad:

We studied the dust storm in Ahmedabad on May 13, 2024, and provided a detailed case study on the dynamics of such events. Dust storms in Ahmedabad are primarily influenced by seasonal variations with higher frequency during pre-monsoon months. The analysis revealed that dust storms contribute to atmospheric instability and disrupting the boundary layer and enhancing aerosol mixing. The backscatter data from the ceilometer highlighted the forcefully changes in the boundary layer dynamics showing increased backscattering due to dust and cloud presence. The backscatter profile shows two layers of obstruction at around 100m and 2500m. The near-surface scattering is due to the presence of dust and the upper-level scattering shows the presence of clouds during the dust storm.

The decrease in long wave flux reflected from the earth during the event was observed by satellites showing the presence of a layer of dust over the place that absorbed the radiation. The storm's impact on temperature and humidity aligns with previous findings, confirming the role of aerosols in atmospheric cooling and humidity increase and strong winds occur. The nocturnal boundary layer collapses and dust particles in the atmosphere reduce surface temperature.

### 4.2.3 Volcanic Eruption in Mount Ruang

The volcanic eruption in Indonesia specifically in Mount Ruang on April 17, 2024, showcased the atmospheric effects of volcanic activity. The eruption released significant amounts of tephra and aerosols into the troposphere, increasing cloud formation and subsequently decreasing surface temperature. The Landsat-9 OLI-II data showed significant before and after changes in vegetation and cloud cover. The temperature variation analysis indicated a notable decrease in temperature following the eruption consistent with historical volcanic events. However, the temperature anomaly cannot directly be related to the event.

The expulsion of $SO_2$ from the volcano was captured by OMI-Aura, which showed a high amount of the gas above Mt. Ruang. $SO_2$ was further carried away by the westward wind to other places. Around 8.5% decrease in the ozone mixing ratio was observed due to the destruction of ozone from the Chlorine emitted by the eruption.

The findings underscore the importance of aerosols in climate modulation, acting as cloud condensation nuclei and influencing longwave radiation flux.

**4.2.4 Instruments and data integration**

We used ground-based instruments like MICROTOPS II sun photometer, Vaisala ceilometer, and AWS monitoring as well as space-based remote sensing such as MODIS (Moderate Resolution Imaging Spectroradiometer), OMI-Aura, Landsat 9, SABER, etc. in studying the atmosphere of the study locations. The integration of observational data in reanalysis models like ERA-5 and remote sensing is accurate and reliable. This study highlights the importance of a multi-instruments approach in atmospheric research particularly in monitoring and analysing rapid and complex changes including natural disasters.

## 4.3 FUTURE SCOPE:

The study of the recent atmospheric events such as the dust storm over Ahmedabad, India, and the volcanic eruption in Mt. Ruang, Indonesia opens up several avenues for further research in this field. The combination of ground-based remote sensing, observational satellite data, and reanalysis data to increase the accuracy of the analysis can be done in a multi-step process. Future advancements in satellite technology will likely provide higher resolution and more accurate data, enabling more detailed analysis of dust storms and volcanic eruptions. This could lead to a better understanding of the fine-scale processes and dynamics involved.

Extending the time scale of the studies to include a more extensive historical dataset can help identify long-term trends and changes in the frequency, intensity, and impacts of dust storms and volcanic eruptions. Comparing the data of a sunphotometer and ceilometer over different regions on a large temporal scale can enhance our understanding of the local atmosphere. Expanding the spatial coverage to include more diverse geographical regions can provide a more global perspective of these events and help identify regional variations and global effects.

Investigating the interactions between dust storms, volcanic eruptions, and climate change can provide insights into how these phenomena influence and are influenced by global climate dynamics. Further research into the health impacts of dust and volcanic ash on human populations, as well as their effects on ecosystems, agriculture, and infrastructure, can inform mitigation and adaptation strategies. Creating and maintaining open-access databases for satellite and reanalysis data can support ongoing and future research efforts, allowing for broader participation and innovation in the field.

# 5 APPENDIX

## 5.2 Codes:

### 5.2.1 Reading '. nc'/ '.nc4' Files:

```
clearvars;
clc;
% Reading and displaying data from .nc extension
ncfile = "filepath.nc";
ncinfo(ncfile);
ncdisp(ncfile);

% Converting time to IST
stet = ncread(ncfile,'time');
T = datetime(1970,1,1) + seconds(stet)+hours(5.5);
% Example of plotting backscattering data from Ceilometer
BSprofile = ncread(ncfile, 'Bs_profile_data');
ht = 10:10:4500;  % Height at a 10 meter interval
pcolor(T,ht,BSprofile);
shading interp; % Interpolating the points
h = colorbar;
ylabel(h, 'Backscattered Photons')
colormap jet;
clim([0 1.5e3]); % Setting the range of color
```

### 5.2.2 Reading '. hdf' Files:

```
f= "Filepath.hdf";
x = hdfinfo(f);
% Code to display variables
for i=1: (no of variables to be read)
    datasets{i,1}=x.Vgroup.Vgroup(2).SDS(i).Name;
end
disp(datasets);
% To extract data from variables
Variable = hdfread(f, 'Variable name'); %Corrected by algorithm
```

### 5.2.3 Reading '. he5' Files:

```
h5file = "filepath.he5";
% Displaying Data from he5 files
h5disp(h5file);
% Reading the data from he5 files
Var1 = h5read(h5file,'/GroupPath/VariableName');
lon = double(h5read(h5file,'/GroupPath/Longitude'));
lat = double(h5read(h5file,'/GroupPath/Latitude'));
% Plotting the data on a worldmap axis
figure;
worldmap('World');
load coastlines;
plotm(coastlat, coastlon);

plt = pcolorm(lat,lon,Var1);
plt.EdgeAlpha=0; % To remove the edges of the grid
shading interp % Interpolating the points
clim([-x x]); % Setting the range of color
colorbar;
```

### 5.2.4 Reading '.csv' Files:

```matlab
aws = "D:\PRL stuff\DATA\AWS\SGSDL22-1013_13052024.csv";
% Reading dataset from .csv file
Dataset = readtable(aws,VariableNamingRule="preserve",NumHeaderLines= 1);
Time = Dataset.("Time Stamp");
AirTemp = Dataset.("Air.Temp (°C)");
relHumidity = Dataset.("Rel.Hum. (%)");
plot(Time,AirTemp, 'LineWidth',2);
yyaxis left;
ylabel(['Temperature in ' char(176) 'C']);
hold on
plot(Time,relHumidity,'LineWidth',2);
yyaxis right;
ylabel('Relative Humidty');
grid on
legend(['Air Temperature in ' char(176) 'C'], 'Relative Humidity','Location',
'northwest');
hold off
```